\documentclass[journal]{IEEEtran}
\usepackage[utf8]{inputenc}
\usepackage[pdftex, pdfborderstyle={/S/U/W 0}]{hyperref}
\usepackage[T1]{fontenc}
\usepackage{afterpage}
\usepackage{listings}
\usepackage{xcolor}
\usepackage{soul}
\usepackage{graphicx}
\graphicspath{{figs/}{figs/metrics/}{figs/dcComparison/} {figs/senders/}
  {figs/ack/} {figs/fastRx2/}}
\usepackage{float}
\usepackage[font=small]{caption}
\usepackage[font=footnotesize]{subcaption}
\usepackage{placeins}
\usepackage{url}
\usepackage{epstopdf}
\usepackage{flushend}
\usepackage{booktabs}
\usepackage{footnote}
\usepackage{multirow}
\makesavenoteenv{table}
\makesavenoteenv{tabular}
\usepackage{tikz}
\usetikzlibrary{patterns}
\usepackage{pgfplots}
\usepackage{dblfloatfix} % fix position of figs when double columns

\usetikzlibrary{positioning,fit, backgrounds, calc}
\usepackage{pgfplots}
\usepackage{microtype}

\usepackage[acronyms,nonumberlist,nopostdot,nomain,nogroupskip]{glossaries}

\usepackage{tabularx}

\usepackage{paralist}

\newacronym{phy}{PHY}{Physical}
\newacronym{mac}{MAC}{Medium Access Control}
\newacronym{ns}{NS}{Network Server}
\newacronym{gw}{GW}{Gateway}
\newacronym{ed}{ED}{End Device}
\newacronym{adr}{ADR}{Adaptive Data Rate}
\newacronym{sf}{SF}{Spreading Factor}
\newacronym{ack}{ACK}{Acknowledgment}
\newacronym{iot}{IoT}{Internet of Things}
\newacronym[plural=LPWANs,firstplural=Low Power Wide Area Networks (LPWANs)]{lpwan}{LPWAN}{Low Power Wide Area Network}
\newacronym{ul}{UL}{uplink}
\newacronym{dl}{DL}{downlink}
\newacronym{qos}{QoS}{Quality of Service}
\newacronym{css}{CSS}{Chirp Spread Spectrum}
\newacronym{dc}{DC}{Duty Cycle}
\newacronym{rx1}{RX1}{first receive window}
\newacronym{rx2}{RX2}{second receive window}
\newacronym{fdgw}{FDGW}{Full Duplex Gateway}
\newacronym{ttn}{TTN}{The Things Network}
\newacronym{ism}{ISM}{Industrial, Scientific, and Medical}
\newacronym{lora}{LoRa}{Long-Range}
\newacronym{toa}{ToA}{time on air}
\newacronym{cpsr}{CPSR}{Confirmed Packet Success Rate}
\newacronym{uu}{UU}{Unconfirmed Uplink PDR}
\newacronym{cu}{CU}{Confirmed Uplink PDR}
\newacronym{cd}{CD}{Confirmed Downlink PDR}
\newacronym{pdr}{PDR}{Packet Delivery Rate}
\newacronym{per}{PER}{Packet Error Rate}
\newacronym{ber}{BER}{Bit Error Rate}
\newacronym{mcu}{MCU}{Micro Controller Unit}
\newacronym{dr}{DR}{Data Rate}
\newacronym{rf}{RF}{radio-frequency}

\title{Energy-Aware Packet Schedulers for Battery-Less LoRaWAN Nodes}

\author{
	\IEEEauthorblockN{Martina Capuzzo\IEEEauthorrefmark{1}, Carmen Delgado\IEEEauthorrefmark{2},
		Jeroen Famaey\IEEEauthorrefmark{3}, Andrea Zanella\IEEEauthorrefmark{4}\IEEEauthorrefmark{1}}\\
	\IEEEauthorblockA{\IEEEauthorrefmark{1}Human Inspired Technologies Research Centre, University of Padova \emph{capuzzom@dei.unipd.it}}\\
	\IEEEauthorblockA{\IEEEauthorrefmark{2}i2CAT Foundation, Barcelona \emph{carmen.delgado@i2cat.net}}\\
	\IEEEauthorblockA{\IEEEauthorrefmark{3}IDLab - Dept. of Computer Science, University of Antwerp - imec \emph{jeroen.famaey@uantwerpen.be}}\\
	\IEEEauthorblockA{\IEEEauthorrefmark{4}Dept. of Information Engineering, University of Padova, \emph{andrea.zanella@unipd.it}}
}

\usepackage{fancyhdr}
\fancypagestyle{FirstPage}{
  \chead{This work has been submitted to the IEEE for possible publication.
    Copyright may be transferred without notice,
    after which this version may no longer be accessible.}
}

\begin{document}

\maketitle

\thispagestyle{FirstPage}

%\pagestyle{empty}
% reduce space before and after eq

%\setlength{\abovedisplayskip}{2pt}
%\setlength{\belowdisplayskip}{2pt}
% space after Figure captions
% \setlength{\belowcaptionskip}{-0.4cm}
%\setlength{\headsep}{0.05in}
%\setlength{\topskip}{-10mm}

%align the bottom of the pages
%\flushbottom

% \setlength{\parskip}{0ex plus0.1ex}
% \addtolength{\skip\footins}{-0.2pc plus 40pt}

%\acmJournal{TOSN}

%\author{Martina Capuzzo}
%\affiliation{%
%  \institution{Human Inspired Technologies Research Centre, University of Padova}
%  \city{Padova}
%  \country{Italy}
%}
%\affiliation{%
%  \institution{Dept. of Information Engineering, University of Padova}
%  \city{Padova}
%  \country{Italy}
%}
%\email{capuzzom@dei.unipd.it}
%
%
%\author{Carmen Delgado}
%\affiliation{%
%  \institution{i2CAT Foundation}
%  \city{Barcelona}
%  \country{Spain}
%}
%\email{carmen.delgado@i2cat.net}
%
%\author{Jeroen Famaey}
%\affiliation{%
%  \institution{IDLab - Dept. of Computer Science, University of Antwerp - imec}
%  \city{Antwerp}
%  \country{Belgium}
%}
%\email{jeroen.famaey@uantwerpen.be}
%
%\author{Andrea Zanella}
%\affiliation{%
%  \institution{Dept. of Information Engineering, University of Padova}
%  \city{Padova}
%  \country{Italy}
%}
%\email{zanella@dei.unipd.it}
%
%
%\pagestyle{empty}

\begin{abstract}

The \gls{iot} enables a wide variety of applications where large sensor networks are deployed in remote areas without power grid access. Thus, the sensor nodes often run on batteries, whose replacement and disposal represent important economical and environmental costs. To realize more sustainable \gls{iot} solutions, it is therefore desirable to adopt battery-less energy-neutral devices that can harvest energy from renewable sources and store it in super-capacitors, whose environmental impact is much lower than that of batteries. To achieve the energetic self-sustainability of such nodes, however, communication and computational processes must be optimized to make the best use of the scarce and volatile energy available. 
In this work, we propose different energy-aware packet scheduling algorithms for battery-less LoRaWAN nodes, and compare them in various simulated scenarios, using actual energy-harvesting measurements taken from a testbed. We show that an energy-aware design  can significantly increase the number of transmitted packets, also lowering the mean time between packet transmissions, though (as predictable) the  gain strongly depends on the harvesting capabilities of the nodes.

\end{abstract}

% \begin{CCSXML}
%<ccs2012>
%<concept>
%<concept_id>10003033.10003039.10003048</concept_id>
%<concept_desc>Networks~Transport protocols</concept_desc>
%<concept_significance>500</concept_significance>
%</concept>
%<concept>
%<concept_id>10003033.10003079.10003081</concept_id>
%<concept_desc>Networks~Network simulations</concept_desc>
%<concept_significance>500</concept_significance>
%</concept>
%<concept>
%<concept_id>10003033.10003039.10003040</concept_id>
%<concept_desc>Sensor Networks~Network protocol design</concept_desc>
%<concept_significance>300</concept_significance>
%</concept>
%</ccs2012>
%\end{CCSXML}
%
%\ccsdesc[500]{Networks~Network simulations}
%\ccsdesc[500]{Sensor Networks}
%% \ccsdesc[300]{Networks~Network protocol design}

%\keywords{Network simulations, ns-3, Internet of Things, Battery-less device,
%  Capacitor, Energy Harvesting, LoRaWAN}

\begin{IEEEkeywords}
	Network simulations, ns-3, Internet of Things, Battery-less device, Capacitor, Energy Harvesting, LoRaWAN
\end{IEEEkeywords}
\maketitle

\section{Introduction}
\label{sec:intro}
\glsresetall
The \gls{iot} paradigm has promoted the development of a number of applications 
where many (smart) objects generate small amounts of data to be delivered to a control station. In many application scenarios, including smart cities, healthcare, and smart agriculture, the devices may not be connected to the power grid and, therefore, need to be powered by batteries. In addition, to prolong their lifetime, they can be equipped with energy harvesting systems. 

Communication typically occurs through long-range wireless data links, which may be provided by 
%In the last years, the rise of the \gls{iot} paradigm promoted the implementation of multiple applications where many (smart) objects are connected to the Internet. Relevant applications include smart cities (e.g., smart lighting systems, public transport monitoring, smart water metering), healthcare (e.g., wearable devices for human parameters monitoring), and smart agriculture (e.g., animal and soil monitoring, precision irrigation).
%These scenarios differ from traditional network applications for the kind of data traffic, which is expected to be sporadic but potentially generated by a very large number of sendor node sources, with limited computational and communication capabilities. Also, the deployment of such devices in wide and remote areas (such as cities and large rural environments) requires to employ long-range wireless communication technologies to minimize the infrastructure. 
\gls{lpwan} technologies ~\cite{centenaro2016long}. In particular, LoRaWAN is a widely deployed solution that supports long coverage ranges (on the order of kilometers in urban scenarios) with extremely low power consumption   \cite{vangelista2015long, world}, and is hence one of the candidate communications technologies for Green \gls{iot}, whose aim is reducing the environmental footprint of \gls{iot} systems~\cite{tahiliani2018green, arshad2017green}. 

For the ecological sustainability of the systems, it is also fundamental to limit the use of batteries (both disposable or rechargeable), since their replacement is costly from a time, economic and environmental perspective. An eco-friendly alternative to batteries is using (super) capacitors to store the energy and power the devices, in combination with energy harvesting systems that can load the capacitor from  renewable sources (e.g., solar, wind, vibrations, ...).

%This approach can boost the application of \gls{iot} solutions in a multitude of scenarios, including the monitoring of wide agricultural plants using solar light~\cite{valente2022lorawan}, of the structural integrity of buildings exploiting vibration energy~\cite{orfei2016vibrations}, of indoor ambient parameters using \gls{rf} energy~\cite{loubet2019lorawan}, of remote areas using the gradient of soil temperature to harvest energy ~\cite{cappelli2021providing, catalan2020experimental}, or for tracking wild animals using solar energy or pizo-electrical harvesters.

%The optimization of LoRaWAN for devices equipped with energy harvesting systems however, is still in its early stage. 
However, the variability of the harvested energy and the small energy density of capacitors can potentially cause an intermittent on/off behavior of the device, affecting its performance and capabilities, including communication. It is therefore important to investigate the behavior of LoRaWAN in nodes with energy harvesting capabilities, in order to assess the feasibility of this solution and configure the network parameters properly. 

In this work, we provide a first analysis of the performance of a battery-less node, equipped with a solar panel for harvesting energy from the environment, and with a LoRAWAN module to transmit and receive data. In our study, we consider real-world energy harvesting measurements collected from a testbed, and a synthetic but realistic node model, whose power consumption profile is also derived from empirical measurements. 
The simulator allows experiments to be carried out in a fully controllable environment, thus enabling different design choices and configurations to be tested with minimal effort, using realistic values for the current drawn by the devices and the energy harvested. Furthermore, the simulator makes it possible to compare different solutions in exactly the same conditions, thus guaranteeing a fair evaluation. %Finally, it permits to evaluate the network performance at scale, and compare many schemes and policies both at the network and device level prior to the actual deployment.
%In this work we employ a network simulator, which is not constrained to particular environmental conditions nor to the specific behavior of devices and/or communication boards, whose hardware implementation can provide variations on the expected behavior. Instead, the usage of the simulator makes it possible to obtain results with more general validity, while allowing to test several devices and configurations with minimum effort, using real measurements for the most important input variables, such as the current drawn by the devices and the harvesting capabilities. Furthermore, the same implementation can be leveraged to evaluate the network performance at scale, and compare many schemes and policies both at the network and device level prior to the actual deployment.

This work is an extension of our previous conference
paper~\cite{capuzzo2021enabling}, which has here been enriched with a more in-depth discussion of scenarios, assumptions, and models, and with new results that highlight the need for energy-aware
approaches, where packet transmission is conditional on the energy level of the device. We evaluate the performance of different packet scheduling approaches in various energy-harvesting conditions. We also investigate the impact of several parameters, such as packet length and capacitor size, by considering a set of performance metrics.

The rest of the paper is organized as follows. In Sec.~\ref{sec:soa} we provide an overview of the literature investigating energy harvesting solutions for \gls{iot}, focusing on LoRaWAN. Sec.~\ref{sec:methods} describes the device's model, the main features of the LoRaWAN technology, and the proposed packet scheduler approaches. The methodology, simulation settings, and results are discussed in Sec.~\ref{sec:results}, while final conclusions are drawn in Sec.~\ref{sec:conclusions}.

\section{State of the Art}
\label{sec:soa}
%
%The adoption of battery-less approaches and energy-harvesting techniques have started gaining interest in the last few years, but many fundamental trade-offs remain to be investigated. For example, the communication could benefit from the robustness provided by message repetitions or the use of a higher transmission power, but this will impact the energy autonomy of the device; conversely, low energy levels may prevent the correct transmission of some packets. These aspects are further complicated when considering the variability of the energy source in nodes with harvested power. 
Works that deal with \gls{iot} networks with battery-less devices and energy harvesting~\cite{fraternali2018pible,   delgado2020battery, sabovic2020energy, gindullina2020energy} mainly address
theoretical analysis (i.e., mathematical modeling) or empirical performance evaluations. 
%Instead, the use of simulations to investigate the interplay between the network state, the system configuration and the device's energy capabilities has not yet been fully exploited. 
As reported in~\cite{peruzzi2020review}, many contributions consider nodes with LoRaWAN communication interface and energy
harvesting capabilities.
In~\cite{mabon2019smaller}, Mabon \textit{et al.} propose the architecture of a general-purpose energy-harvesting sensor node powered with solar panels. Instead, other authors focus on more specific application scenarios, or on the energy-harvesting aspects:
in~\cite{polonelli2018accurate, boccadoro2019quakesense} the authors integrate solar panels to monitor buildings' structural health and detect earthquakes, respectively. Solar panels are also used as a source of
energy in \cite{addabbo2021solar, sadowski2020wireless}. The first work targets a smart city scenario, with a sensor measuring particulate matter concentration, while the second considers a smart agriculture scenario. In
\cite{moid2020towards}, the authors investigate the usage of kinetic energy harvesting to convert motion/vibration energy into electrical energy to power \gls{iot} devices and show the efficiency of the proposed solution using a prototype node. In~\cite{finnegan2020exploring} Finnegan \textit{et al.}
explore the feasibility of a LoRaWAN sensor powered with \gls{rf} ambient resources  and identify some requirements on the hardware design and environmental conditions to permit the proper functioning of the system. In particular, the availability of wireless energy (which depends on the considered scenario) and the distance between  transmitter and receiver are identified as critical factors. Similarly, in~\cite{loubet2019lorawan}, Loubet
\textit{et al.} describe a prototype implementation of a LoRaWAN monitoring network where devices are powered with energy harvested from \gls{rf} sources and propose to control the periodicity of the measurement and data transmission based on the availability of \gls{rf} resources. In~\cite{orfei2016vibrations} Orfei \textit{et al.} describe a system to monitor the asphalt of a bridge through LoRaWAN sensors powered by the bridge vibrations. Analogously, in~\cite{cappelli2021providing}, the authors present the prototype of a LoRaWAN node powered with thermo-electric generators, with experiments showing good results. In~\cite{gleonec2021energy}, Gleonec \textit{et al.} propose a method to adapt the quality of service of a LoRaWAN node to the energy capabilities while employing multi-source energy harvesting. The power management system solution is then validated with off-the-shelf devices in an industrial setting and shows an increase in the number of transmitted messages of 51\%.

% \cite{casals2017modeling} model of LoRaWAN energy performance based on real
% experiments. It was observed that, according to the application requirements,
% the estimated battery lifetime can vary from 1 to 6 years. Also, communication
% settings, such as the usage of confirmed/unconfirmed traffic, have significant
% impact on the energy performance, and that the energy-cost per delivered bit is
% much lower when using maximum-size frames, meaning that it is better to
% accumulate readings and sending them sporadically, then sent frequent shorter
% packets. OPPOSITE TO OUR SCENARIO (DO SOME CONSIDERATIONS)

While these works adopt an empirical approach, based on prototyping and actual testing in the field, other contributions propose theoretical
approaches to gain a better understanding of more fundamental trade-offs, but typically make some simplistic assumptions (capacitor size, fixed energy harvesting rate). In~\cite{delgado2020battery} Delgado
\textit{et al.} employ Markov chains to model the intermittent behavior of a single LoRaWAN device with energy harvesting, including the effects of parasitic resistances of the capacitor, evaluating feasibility and performance when using \gls{ul} and \gls{dl} traffic, and validating it with a C++ simulator. The work presented in~\cite{sabovic2020energy} considers a battery-less capacitor-based LoRaWAN device, and investigates proper scheduling of sensing and transmission
tasks, comparing a turn-off and a sleep-based approach. Furthermore, it studies the optimal turn-on voltage threshold that allows the node to complete such tasks for a given setting (capacitor size, energy harvesting rate), and validates the results using an environment emulator connected to a real device.

Despite this growing interest in battery-less nodes with energy harvesting, many fundamental trade-offs remain to be investigated. For example, when message repetition can be used to increase robustness without rapidly depleting all the node's energy, or whether it is more convenient to increase the transmission power to achieve higher bitrates and, hence, reduce the packet transmission time, or adopt  more robust low bitrate modulations; how packet transmissions should be scheduled according to the current energy harvesting rate, and so on. 

With respect to the state-of-the-art, in this paper, we focus on a single-node scenario and analyze the impact of different parameters on the device's performance. We extend the results shown in~\cite{capuzzo2021enabling} by comparing more energy-aware packet scheduling approaches and employing various energy harvesting traces obtained from real measurements. To evaluate the feasibility of battery-less LoRaWAN nodes with energy harvesting, we leverage ns-3 simulations, with the implementation we previously described in~\cite{capuzzo2021ns}. This allows us to investigate the interplay between the network state, the system configuration, and the device's energy capabilities, which has not yet been fully analyzed and understood. 

\section{Modeling of a battery-less LoRaWAN device and scheduling approaches}
\label{sec:methods}
In this section, we illustrate the model of a battery-less device, equipped with a supercapacitor and LoRaWAN technology, which will also be briefly presented.  In addition, we introduce
some packet scheduling approaches that implement different energy-aware algorithms to determine when to transmit sensor data. Their performance will be evaluated in Sec.~\ref{sec:results}.

\subsection{Battery-less IoT devices with energy harvesting}
\label{sec:model}
\begin{figure}[t]
  \centering
  \begin{picture}(100,120)
    \put(-50,20){\includegraphics[width=0.85\columnwidth]{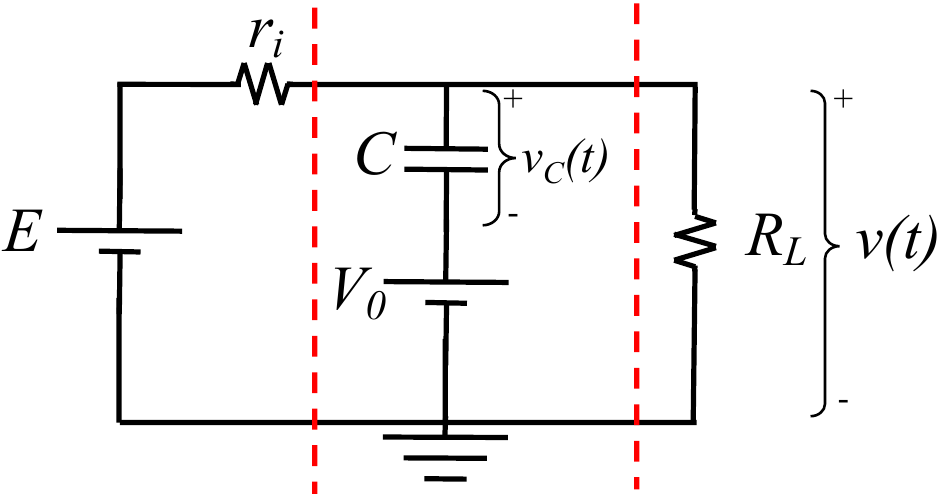}}
    \put(-30,10){Harvester} \put(30,10){Capacitor} \put(90,10){Load: MCU,
      Radio,...}
  \end{picture}
  \caption{Electrical circuit model of a battery-less IoT device~\cite{delgado2020battery}.}
  \label{fig:circuit}
    \vspace{-0.4cm}
\end{figure}
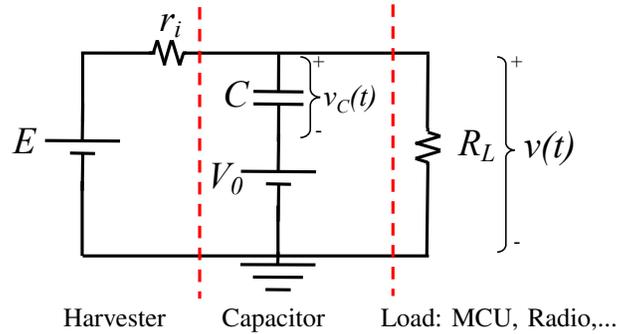
Following the approach proposed  in~\cite{delgado2020battery, capuzzo2021ns}, we model the device as an equivalent electrical circuit with three parts accounting for, respectively: (i) the harvester, (ii) the capacitor, and (iii) the load. Fig.~\ref{fig:circuit} depicts the
corresponding circuit.

\begin{enumerate}[(i)]
\item \emph{The harvester:} it is the only energy source in the system. The harvester is modeled as an ideal constant voltage source (denoted by $E$) with a time-dependent series resistance $r_i(t)$. In general, the resistance $r_i(t)$ changes in time to reflect the fluctuations of the energy harvesting process. The maximum power that can be actually produced by the harvester at time $t$ is hence given by
\begin{equation}
P_{h}(t) = \frac{E^2}{r_i(t)}\,.
\end{equation}
In the following, we will
consider both an ideal harvesting source, characterized by a constant harvesting
power, and realistic energy sources, whose harvested power is taken from empirical traces obtained in a testbed, using solar panels (see Sec.~\ref{sec:ehtraces} for details).
\item \emph{The load:} it is modeled as a resistance, $R_L(s)$, that represents all the equivalent load of all the system's components that consume energy in a certain operating state $s$ of the device (e.g., idling, transmission, processing). 
\item \emph{The capacitor:} it stores the energy provided by the harvester and releases it to the load when required. The voltage of the capacitor can be represented by $(V_0(s), v_C(t))$, where $V_0(s)$ and $v_c(t)$ are the capacitor voltage when entering a certain state $s$, and after $t$ seconds spent in that state, respectively. As depicted in Fig.~\ref{fig:circuit}, $V_0(s)$ is modeled as an ideal voltage source, while $v_C(t)$ is the voltage over time of an ideal capacitor (with $v_C(0)=0$). The voltage provided by the capacitor to the load after $t$ seconds in state $s$ is then given by 
\begin{equation}
    v(t, s)= E \frac{R_{eq}(s, t)}{r_i(t)}\big(1 - e^{-\frac{t}{R_{eq}(s, t)C}}\big) + V_0(s) e^{-\frac{t}{R_{eq}(s, t)C}},
    \label{eq:vc}
\end{equation}
where $C$ is the capacitance of the capacitor, and
\begin{equation}
    R_{eq}(s, t) = \frac{R_L(s)r_i(t)}{R_L(s) + r_i(t)}
\end{equation}
is the equivalent resistance seen by the capacitor.
\end{enumerate}

To be operational, real devices need a voltage above a certain threshold, which we
denote by $V_{th\_low}$. When the capacitor's voltage drops below this value,
the device switches off, and it cannot perform any operation. In this state, the current consumption is minimal% (5.5~$\mu$A, due to the circuitry~\cite{delgado2020battery})
, and most of the harvested energy will hence be used to recharge the capacitor. When the stored energy exceeds the
voltage threshold $V_{th\_high}$ (larger than $V_{th\_low}$), the device switches back to the active state. Note that, this wake-up operation also consumes some energy.

\subsection{LoRaWAN technology}
\label{sec:lorawan}
%
% \begin{figure}[t]
%   \centering
%   \includegraphics[width=0.75\columnwidth]{infrastructure.png}
%   \caption{LoRaWAN infrastructure. }
%   \label{fig:infrastructure}
% \end{figure}
%
LoRaWAN~\cite{lorawan} is based on the proprietary LoRa modulation, which leverages the \gls{css} modulation scheme. Robustness is increased by a time-spreading coding scheme defined by the \gls{sf} parameter, which takes integer values from 7 to 12. For a given channel bandwidth, the \gls{sf} is inversely proportional to the data rate: the higher the \gls{sf}, the more robust the transmission (and, hence, the longer the coverage range, but the lower the data-rate and the packet transmission time.
% Another important feature of LoRa is that signals modulated with different \glspl{sf} are almost orthogonal: therefore, two signals can be simultaneously decoded even if they overlap in time and frequency, provided that they satisfy some mild conditions on the relative received powers.
% in~\cite{croce2018impact}. Similarly, when multiple signals modulated with the
% same \gls{sf} overlap at the receiver, one signal can be correctly demodulated
% if its power is at least $\rho$~dB higher than the sum of the powers of the
% interferers (capture effect). The co-channel rejection threshold $\rho$ is
% specified in LoRa chip datasheets~\cite{sx1301, sx1272}, and can be as large as
% 25 dB for \gls{sf} 12, as estimated in~\cite{croce2018impact}.

%\glsreset{ns, gw, ed}
The LoRaWAN standard also defines the topology of the network. It considers a star-of-stars topology with three types of devices:
\textit{(i) \glspl{ed}}, i.e., peripheral nodes (usually sensors or actuators) that communicate using the LoRa modulation with the \glspl{gw};
\textit{(ii) \glspl{gw}}, that relay the messages coming from the \glspl{ed} through the LoRa interface to the \acrlong{ns} using a reliable IP connection, and vice-versa; the \textit{(iii) \gls{ns}}, i.e., the central network controller that is the endpoint of all \glspl{ed} communications, and is in charge for setting the network parameters.

%Fig.~\ref{fig:infrastructure} illustrates an example of a LoRaWAN network, where dashed lines represent LoRa links, while IP connections are shown as solid lines.

% % Some words about the gw...
% When the \glspl{ed} transmit, the LoRa packets are collected by all the \glspl{gw} in their coverage range and forwarded to the \gls{ns}, which discards duplicates and chooses the best \glspl{gw} for the return transmissions to the \glspl{ed}. It is worth noting that the \glspl{gw} do not support full-duplex
% transmission: to send a \gls{dl} packet (from the \gls{ns} to an \gls{ed}), they have to interrupt any ongoing \gls{ul} reception.
% Classes of devices
The LoRaWAN specifications define three classes of \glspl{ed}, which differ
in terms of energy-saving capabilities and reception availability. In this work,
we focus on Class A devices, which stay in sleep mode most of the time in order
to minimize energy consumption. 

Fig.~\ref{fig:cycle} represents the sequence
of operations performed by a Class A device. The \gls{ed} can wake up at anytime
(time $t_0$) to transmit its data. Then, it enters into the idle state, waking up after one and two seconds to check for incoming (reply) packets. More specifically, at time $t_2=t_1+1$~s, it opens the \gls{rx1}. If no incoming signal is detected, the \gls{ed} switches back to the idle state and, at time $t_3 = t_1+2$~s, it opens the \gls{rx2} for another short interval. After the reception phase is over, the \gls{ed} enters a sleep state, where it remains till the next packet transmission. In the following,
we will indicate as \emph{transmission cycle} the period between two consecutive
packet transmissions, including all the intermediate phases.

\begin{figure}[t]
  \centering
 % \hspace{-2cm}
  \begin{subfigure}{\columnwidth}
    {
    \centering
   \resizebox{0.9\columnwidth}{!}{
      }
    }
  \end{subfigure}
  \caption{Example of transmission cycle for a Class A LoRaWAN node.}
  \label{fig:cycle}
\end{figure}
The specifications define two types of messages: confirmed and unconfirmed. In
the first case, each \gls{ul} transmission is expected to be confirmed by the \gls{ns} through the transmission of an
\gls{ack} packet, which must be received  by the \gls{ed} in \gls{rx1} or \gls{rx2}. If the \gls{ack} is
not received, the \gls{ed} can re-transmit the packet. Unconfirmed messages, instead, do not require any \gls{ack} and are generally
transmitted only once.

% Channels
LoRaWAN operates in the unlicensed \gls{ism} frequency bands. The relevant
regulations define frequency bands, power and \gls{dc} restrictions to be
applied.
% In Table~\ref{tab:channels}, we report the default values for the
% European region as recommended by the LoRaWAN specifications; although the
% use of additional channels on other sub-bands is permitted, in this work we
% focus on the minimum frequency set mandated by the specifications.
In the European region, LoRaWAN~\cite{lorawan} defines the use of three 125~kHz
wide channels, centered at 868.1~MHz, 868.3~MHz, 868.5~MHz, which are shared between
\gls{ul} and \gls{dl} transmissions, and must \textit{collectively} respect a 1\%
\gls{dc} constraint. A fourth 125~kHz channel centered at 869.525~MHz is used for \gls{dl} communication only and is subject to a \gls{dc} constraint of 10\%~\cite{rec7003e}. As an example, in
Tab.~\ref{tab:dcexamples} we report some cases of packet time-on-air ($t_{packet}$) and minimum interval between transmissions, as dictated by the \gls{dc} constraint, for different packet sizes and \glspl{sf}.

\begin{table}[t]
  \centering
  \caption{Minimum transmission interval allowed by a duty cycle of 1\% for
    different packet sizes and \glspl{sf} for \gls{ul} transmissions, with MAC+PHY overhead of 13 bytes.}
  % \begin{tabularx}{\columnwidth}{llX}
   \begin{tabular}{c c c c c}
     \toprule
     & \multicolumn{2}{c}{SF 7} & \multicolumn{2}{c}{SF 8} \\
    \toprule
     Payload & $t_{packet}$  & Min interval & $t_{packet}$ & Min interval \\
    \midrule
    0 B & 46.34 ms & 4.34 s & 82.43 ms & 8.24 s \\
    5 B & 51.46 ms & 5.15 s & 92.67 ms & 9.27 s\\
    50 B & 118.02 ms & 11.80 s & 215.55 ms & 21.56 s \\
    100 B & 189.70 ms & 18.97 s & 338.43 ms & 33.84 s\\
    \bottomrule
  \end{tabular}
  \label{tab:dcexamples}
  \vspace{-0.4cm}
\end{table}

% Default configuration of RX windows
For Class A devices, the standard requires \gls{rx1} to be opened in the same frequency channel and with the same \gls{sf} used for the \gls{ul} communication. \gls{rx2}, instead, is opened on the dedicated 869.525~MHz channel and with \gls{sf}~12, by default, in order to maximize the robustness of the communication. These configurations, however, can be modified by the \gls{ns}.

\subsection{Packet scheduling approaches}
\label{sec:senders_metrics}
To evaluate the feasibility of battery-less LoRaWAN devices powered with energy harvesting systems, we investigate the design of different packet scheduling approaches that take the device's energy level into account and make the best use of the available energy resources. The schedulers can be categorized into three groups: (i) benchmark (see later); (ii) energy-aware, which 
considers only the current energy level of the nodes; (iii) energy-modeling, which also predicts the energy available after the transmission. We take as reference an \gls{iot} application that generates a new packet every $I$ seconds, overwriting any previous packet still in the buffer. 

As benchmarks we consider two extreme cases. On the one hand, we define the \textbf{Unaware Sender (US)}, which generates packets with a period $I$, independently of the device's energy level. This energy unaware approach will be used as the baseline for comparison with the energy-aware approaches and provides a lower performance bound.
On the opposite end, we define the \textbf{Optimal Sender (OS)} that assumes perfect non-causal knowledge of the harvested power during the whole cycle, so that packet transmission occurs as soon as the energy stored in the capacitor is sufficient to sustain the transmission cycle, taking into account the additional energy that will be harvested in the meanwhile. While this scheduler is not realistic, its performance represents an upper bound to any feasible scheduler in the considered scenario.

As energy-aware scheduler, we consider the \textbf{Energy-Aware Sender with Fixed Threshold (FS)}: it schedules a transmission only if the voltage level of the capacitor is above a fixed threshold $V_{th\_FS}$.

The class of energy-modeling schedulers, finally, consists of algorithms that approximate the OS, but without assuming non-causal knowledge of the harvested energy. Therefore, these schedulers allow for packet transmission only if the capacitor energy is sufficient to successfully complete the transmission cycle, preventing the device from switching off. The threshold is computed taking into account the packet size, \gls{sf}, and \gls{ack} reception (when required).
Then, depending on the mechanism used to predict the harvesting power $\hat P_h(t)$ during the transmission cycle, we define the following energy-modeling schedulers:

\textbf{Conservative Sender (CS)}: it conservatively assumes no energy harvesting ($\hat P_h(t)=0$).

\textbf{Simple moving Average Sender (AS):} the harvesting energy is assumed to equal the \emph{mean} harvested energy over the last $x$ seconds ($\hat P_h(t)=\bar{P}_h(t) = \frac{1}{x}\int_{t-x}^t P_h(u)\mathrm{d}u $). In the following, we will indicate this solution as AS-$x$.

\textbf{Minimum harvesting Sender (MinS):} the harvesting power is assumed to be equal to the \emph{minimum} value during the previous $x$ seconds ($\hat P_h(t)  =\min\{P_h(u); u\in [t-x,t)\}$). Note that this approach is more conservative than AS. In the following, we will consider $x=5$~s.

\textbf{Average and Variance Estimation Sender (AVES):} this approach computes the expected harvesting power by considering the moving-average estimates of mean and variance over time, similarly to what done in~\cite{jacobson1988congestion}.
At time $i$, the computation considers the average $\bar{P}_h(i)=\frac{1}{x}\int_{i-x}^i P_h(u)\mathrm{d}u $ of the harvested power over the previous $x$ seconds. The mean $A_i$ and the deviation $D_i$ are then computed as
%
% \item \textbf{Average and Variance Estimation Sender (AVES):} this approach
%   computes the expected harvested power by considering average value and
%   variance over the last period, similarly to what done
%   in~\cite{jacobson1988congestion}. Each time, the computation takes into
%   account the current measure of the average at time $i$, $M^i$ and its
%   difference with the same value at time $i-1$. Then, with $g$ a weighting
%   coefficient, $A$ the average harvested power and $D$ the mean deviation, we
%   have
  %
  \begin{equation}
    \begin{split}
      A_i &= g\bar{P}_h(i) + (1 - g) \cdot A_{i-1}, \\
      D_i &= g \cdot |\bar{P}_h(i) - A_{i}| + (1-g)\cdot D_{i-1},
     \end{split}
  \end{equation}
  where $g$ is a weighting factor. The estimated harvesting power $\hat{P}_h(t)$ is finally computed as
  \begin{equation}
    \label{eq:hphat}
    \hat{P}_h(t) = \max(A_{t}-D_{t}, 0).
  \end{equation}
In the results presented in the following, we assume $x = I$, and we take $g = 0.1$. Preliminary investigations with the same settings showed similar outcomes for other values of $g$.

\begin{table*}[t]
  \centering
  \caption{Main features of harvested power traces.}
\begin{tabular}{lllcc}
  \toprule
  Trace & Day of collection & Position & Mean power [mW] & Reference\\
  \midrule
  Trace A & sunny day in September 2020 & West-facing windowsill & 7.2$\pm$8.2 & \small{self-collected in Antwerp (BE)}\\
  Trace B & cloudy day in September 2020 & West-facing windowsill & 4.0$\pm$2.0 & \small{self-collected in Antwerp (BE)}\\
  Trace C & 12th June, 2010 & North-facing windowsill & 27.5$\pm$8.1 & \cite{gorlatova2011networking}, Setup C \\
  Trace D & 12th December, 2009 & North-facing windowsill & 3.8$\pm$3.3 & \cite{gorlatova2011networking}, Setup C \\
  \bottomrule
\end{tabular}
\label{tab:power}
\vspace{-0.2cm}
\end{table*}

% Note on the DC
Packets that cannot be transmitted because of \gls{dc} constraints are
immediately dropped. This assumption results in different behaviors between US
and the other approaches because when setting $I$ to be lower than the minimum
silent time imposed by the \gls{dc}, an \gls{ed} whose transmission has been prevented by the \gls{dc} will wait $I$ more seconds before the generation of
the next packet under the US algorithm, while with the energy-aware algorithms, once allowed by the \gls{dc} constraint the packet will be generated and transmitted as soon as the energy level permits.

\section{Results}
\label{sec:results}
In this section, we describe the settings employed in our simulations and the
performance metrics that we leverage to evaluate the different packet scheduling approaches. Then, we showcase and discuss the simulation results.

\subsection{Energy harvesting traces}
\label{sec:ehtraces}

\begin{figure}[t]
  \centering
  \includegraphics[width=0.7\columnwidth]{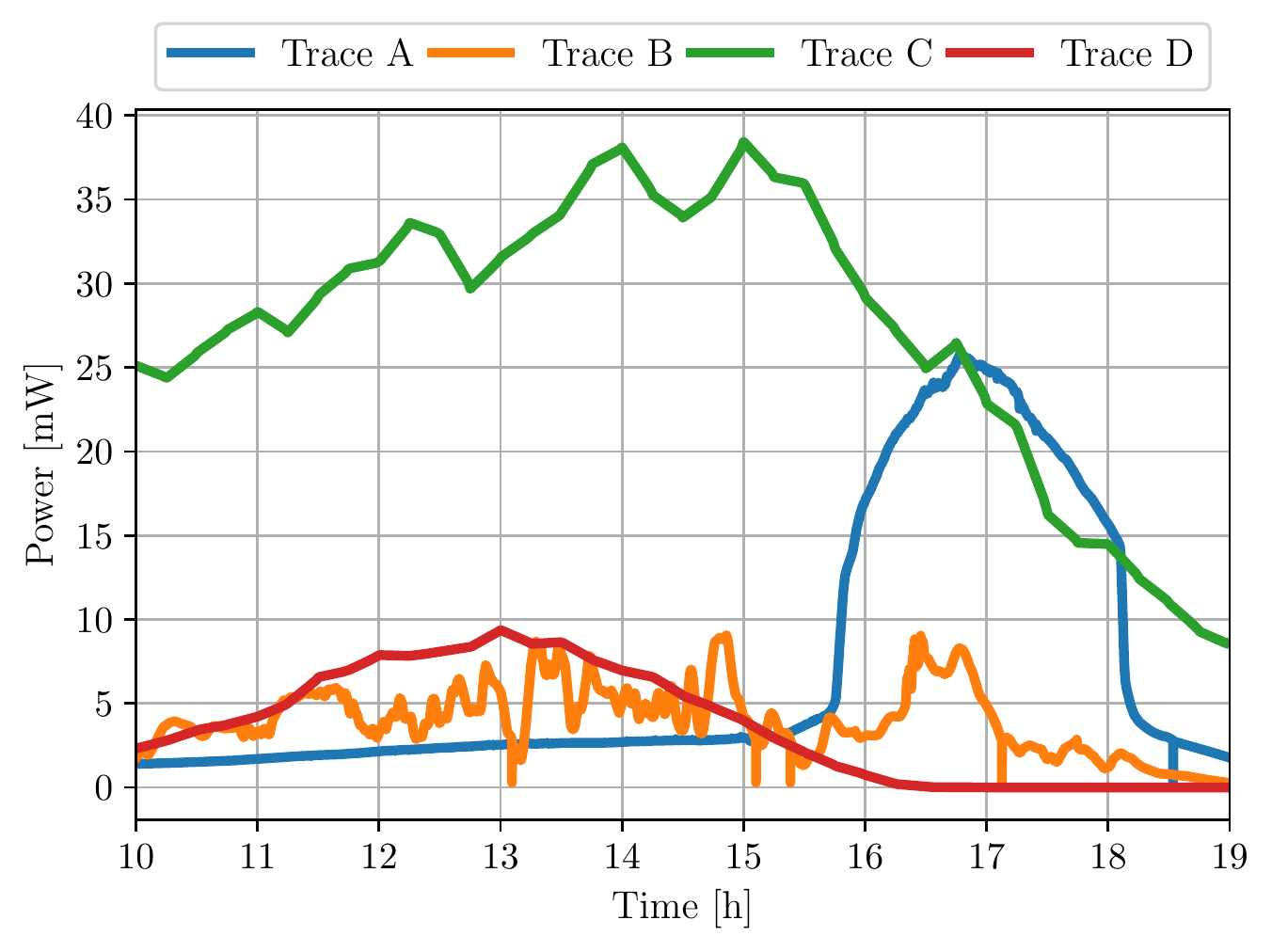}
  \caption{Harvested power $P_h$ for different traces.}
  \label{fig:power}
  \vspace{-0.5cm}
\end{figure}

In the following analysis, we test the device's performance with different packet scheduling approaches, considering different kinds of harvesting sources, namely: \textit{(i)} an ideal energy harvesting source providing constant harvesting power, \textit{(ii)} several empirical traces of power harvested by solar panels in indoor environments. Specifically, we consider 4 power traces collected from 10:00 to 19:00 in different periods of the year, locations and employing different hardware, as detailed in~Tab.~\ref{tab:power}. All measurements have been collected by employing a 6-cells mono-crystalline (4x2~cm$^2$) solar panel. As it can be seen in Fig.~\ref{fig:power}, the harvested power varies significantly with the scenarios, making it possible to analyze the system performance in heterogeneous
conditions. 
\subsection{Simulation settings}
For our simulations, we leverage the \texttt{lorawan} ns-3 module and the
capacitor implementation described in~\cite{capuzzo2021ns}, extended to evaluate
different packet schedulers. We simulate a single-gateway single-\gls{ed}
network, with the \gls{ed} transmitting packets with different data payloads
(PL) using \gls{sf}~7. Note that, since a single device is employed, varying the
\gls{sf} would only yield a rescaling of the results, without changing the
considerations that can be drawn from the analysis. Furthermore, once the
configuration is set, the simulation is deterministic, thus the confidence
intervals are not represented in the plots.

The \gls{ed} is provided with a capacitor that is initially empty and then gets
charged by the harvesting process. The duration of the simulations was set to
9~hours, based on the available energy harvesting traces. The device switches
off if the capacitor's voltage falls below $V_{th\_low}=1.8$~V, while the
reactivation threshold is set to $V_{th\_high}=3$~V. These values that are in line
with off-the-shelf LoRa devices~\cite{delgado2020battery}.
%Tab.~\ref{tab:current} reports the current consumption for the different device's states.
The minimum interval between the generation of consecutive
packets is set to $I=4$~seconds. Note that, for $\mathrm{PL} \geq 5$~B, the
maximum throughput is actually limited by the \gls{dc} constraint. In the
results shown below, the voltage threshold for the FS algorithm is
set to $V_{th\_FS} = 1.82$~V, slightly above the $V_{th\_low}$ threshold to
allow as many transmissions as possible.
%

% \begin{table}[tb]
%   \centering
%   \caption{Current consumption in the different states~\cite{delgado2020battery}.}
%   \begin{tabular}{llll}
%     \toprule
%     State & MCU & Radio current & Total current \\
%     \midrule
%     Off &  Standby & 0 & 5.5 $\mu$A \\
%     Turn On & Active & - & 15 mA \\
%     Sleep & Active & 1 $\mu$A & 5.6 $\mu$A \\
%     Tx & Active & 28 mA & 28.011 mA \\
%     Idle & Standby & 1.5 $\mu$A & 7 $\mu$A\\
%     Standby & Standby & 10.5 mA & 10.5055 mA\\
%     Rx & Active & 11 mA & 11.011 mA \\
%     \bottomrule
%   \end{tabular}
%   \label{tab:current}
%     \vspace{-0.5cm}
% \end{table}

\subsection{Performance metrics}
To compare the different scheduling algorithms, we will employ the following
metrics:
\begin{itemize}
\item Number of \gls{ul} packets successfully transmitted by the \gls{ed}. Note
  that it may be possible that a packet is successfully transmitted, but the
  \gls{ed} is not able to complete the cycle because of a low voltage value. In
  this case, the packet transmission is successful, but the device will switch
  off, possibly preventing future transmissions if not able to recharge on time.
\item Percentage of \gls{ul} packets that are successfully transmitted by the
  \gls{ed} with respect to the maximum possible one, considering the limitation
  imposed by the \gls{dc}. This represents the \emph{efficiency} of the system,
  and is computed as
  \begin{equation}
    \epsilon = \frac{p_{sent}}{p_{max}}\cdot 100,
  \end{equation}
  where $p_{sent}$ is the number of packets sent by the \gls{ed} during the
  whole simulation, and $p_{max}$ is the maximum number of packets allowed by
  the regulations in the sub-band with 1\% \gls{dc}, i.e., $p_{max} =
  \frac{simulation\;time}{100 \cdot t_{packet}}$, where $t_{packet}$ is the time
  on-air of a packet transmission.

\item Fraction of time spent in the different states, namely: operational state
  (ON - sleep, transmission, reception), non-operational (OFF), or initial charging phase (Charging), i.e., when
  charging the capacitor to $V_{high}$ for the first time.
\item Mean time between consecutive transmissions, neglecting the initial
  charging phase. Note that, in the best case (i.e., no energy constraints),
  this is lower bounded by the \gls{dc} limitation.
\end{itemize}

\subsection{Results and discussion}
\subsubsection{Importance of energy-awareness}

\begin{figure}[]
  \centering
  \includegraphics[width=0.9\columnwidth]{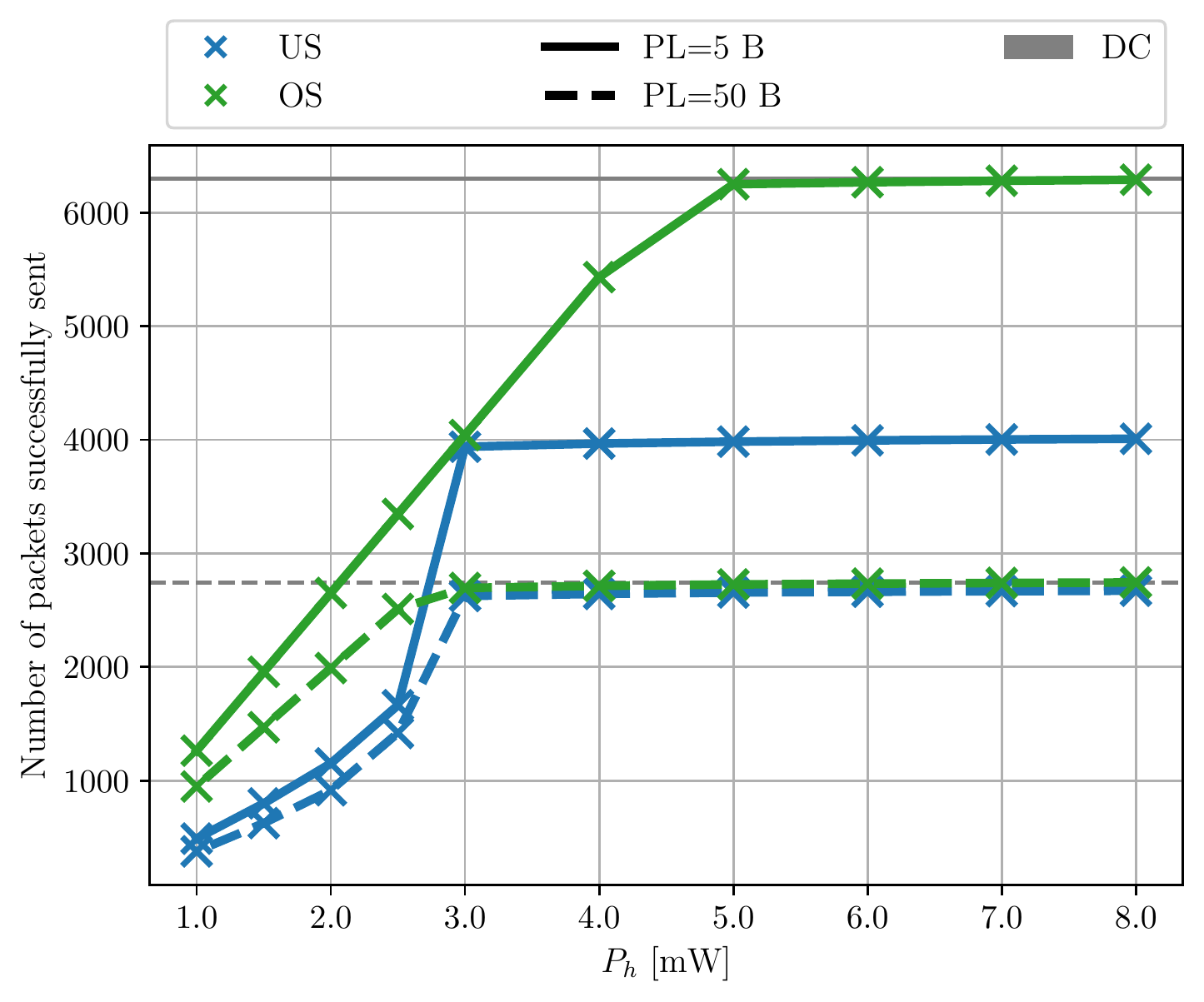}
  \caption{Performance for different configurations, with constant $P_h$ and
    C=100~mF.}
  \label{fig:constant}
    \vspace{-0.5cm}
\end{figure}

We begin our discussion by assessing the maximum gain in number of sent packets that can be obtained moving from an energy unaware algorithm (US) to the ideal energy-aware scheduling approach (OS). Also, we show the impact of the
packet size, with the assumption of constant harvesting power $P_h$.  Fig.~\ref{fig:constant} shows the number of transmitted packets when varying the harvesting power $P_h$. As expected, all schedulers transmit more packets for higher values of $P_h$, until the maximum value imposed by the \gls{dc} limitations is reached. Looking at Tab.~\ref{tab:dcexamples}, indeed, the maximum number of packets that can be transmitted in the considered period (9~hours) is 6296 for PL=5~B, and 2745 for PL=50~B, which correspond to those achieved by the schedulers for high harvesting power. (Actually, UL does not reach the maximum theoretical performance for $\mathrm{PL}=5$~B because packets generated during the silent period imposed by the \gls{dc} are immediately dropped, while OS adjusts the packet generation time accordingly.)

In general, OS always outperforms US, in particular for short  packets (solid lines). Instead, for long packets (dashed lines) and sufficiently high $P_h$, the performance gap vanishes, because the \gls{ed} can recover for the energy spent in the transmission of long packets during the also long silent periods imposed by the \gls{dc}. 

Although not reported here for space constraints, we observed that the capacitance value  has a limited impact on the performance of these benchmark algorithms, with a slight increase of OS's performance when decreasing the capacitance to 20~mF, as a result of the shorter recharging time. The difference in the number of packets between the US and the OS is further discussed in Sec.~\ref{sec:perfEval}.

% % Comparison of senders
\begin{figure*}[]
  \centering
  \begin{subfigure}{0.75\columnwidth}
    {
      \includegraphics[width=\linewidth]{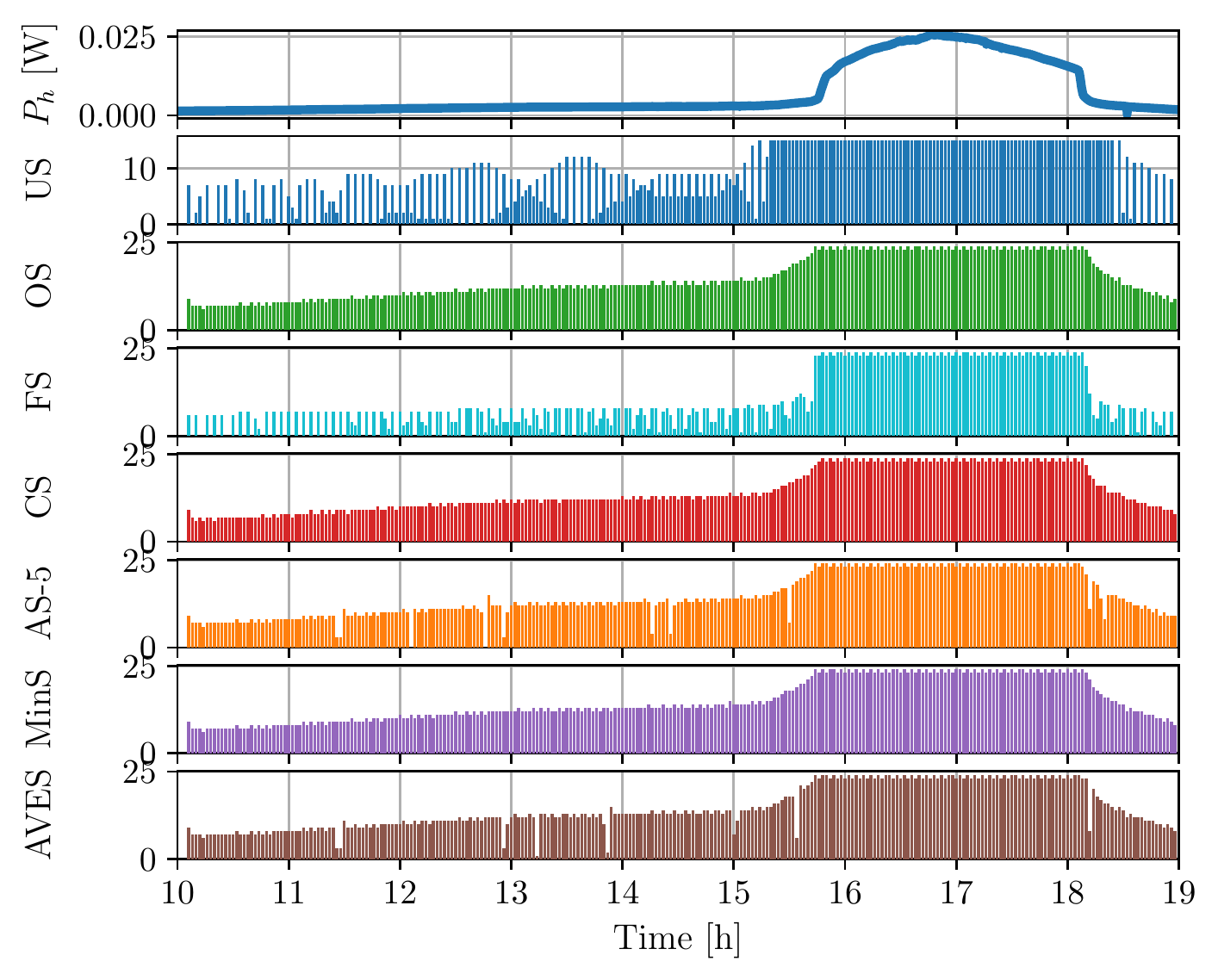}
    }
    \vspace{-0.4cm}
    \caption{Number of transmitted packets for different algorithms, C=20~mF.}
    \label{fig:sendertime20}
    %\vspace{0.7cm}
  \end{subfigure}
  \hspace{0.7cm}
  \begin{subfigure}{0.75\columnwidth}
    {
      \includegraphics[width=\linewidth]{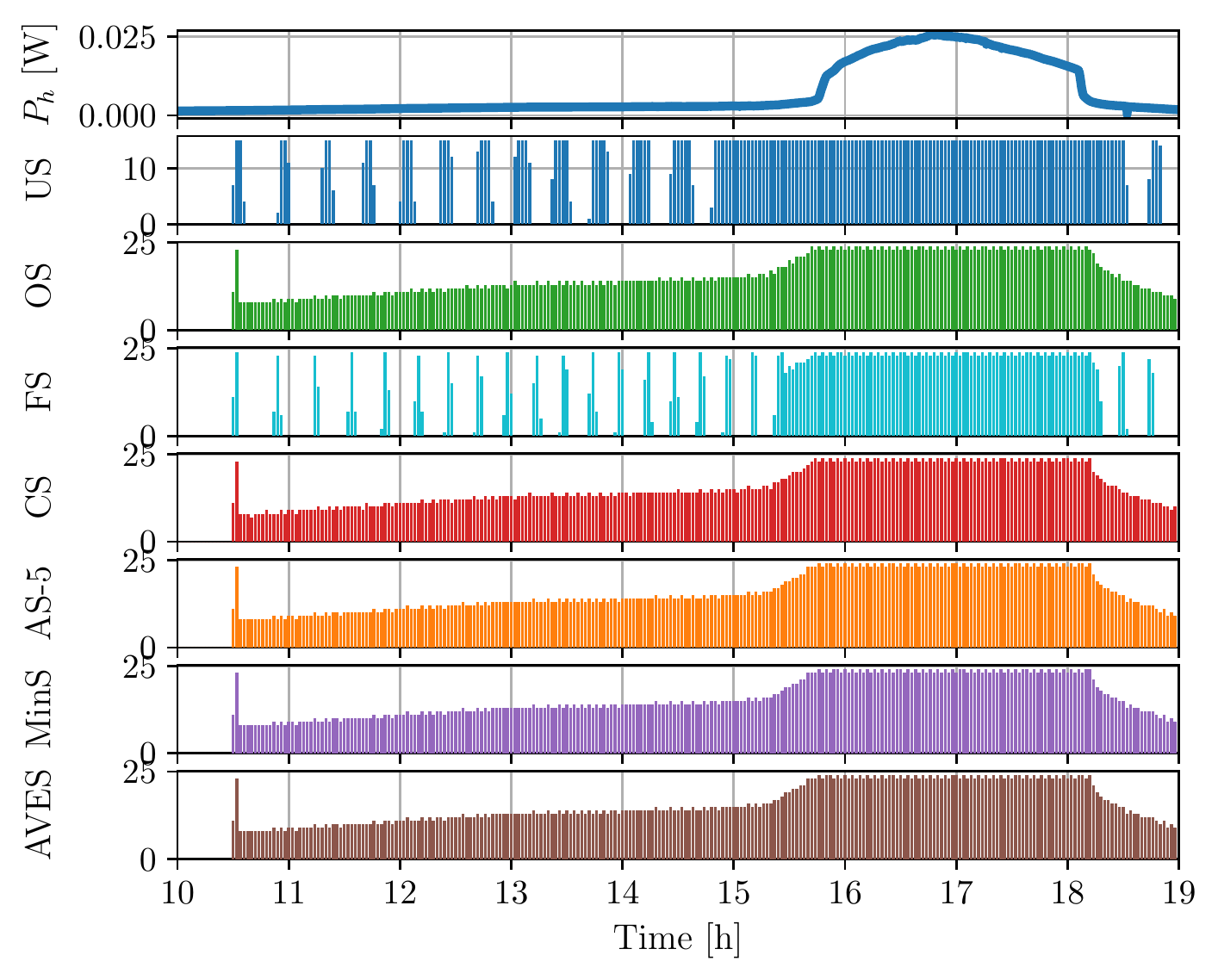}
    }
    \vspace{-0.4cm}
    \caption{Number of transmitted packets for different algorithms, C=100~mF.}
    \label{fig:sendertime100}
    %\vspace{0.7cm}
  \end{subfigure}
  \begin{subfigure}{0.6\columnwidth}
    {
      \includegraphics[width=\linewidth]{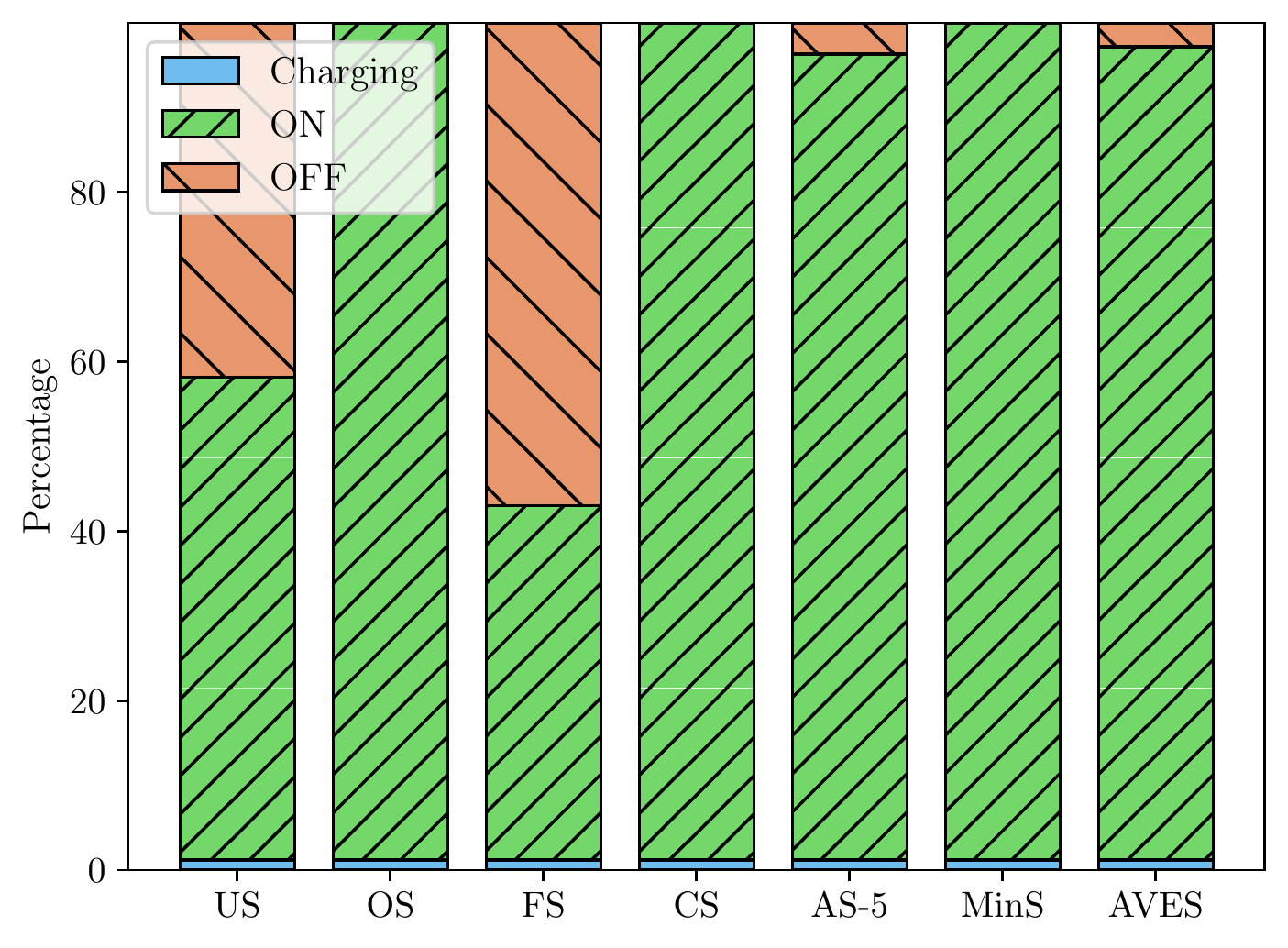}
    }
    \vspace{-0.5cm}
    \caption{Fraction of time spent in different states, C=20~mF.}
    \label{fig:onoff20}
  \end{subfigure}
  \hspace{2.2cm}
  \begin{subfigure}{0.6\columnwidth}
    {
      \includegraphics[width=\linewidth]{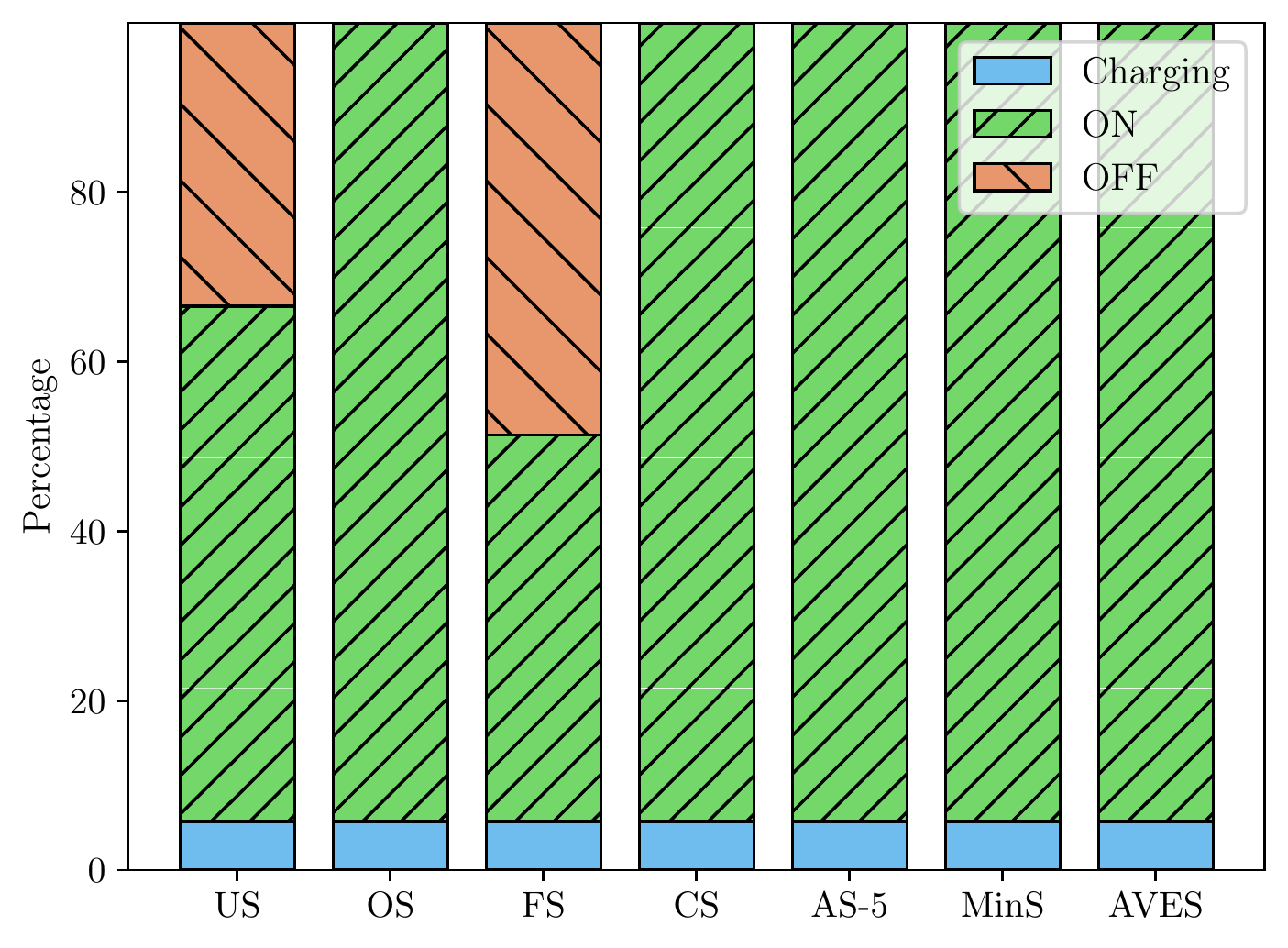}
    }
    \vspace{-0.5cm}
    \caption{Fraction of time spent in different states, C=100~mF.}
    \label{fig:onoff100}
  \end{subfigure}
  \caption{Comparison of scheduling behaviors for different capacitances, using trace A.}
  \label{fig:senderstime}
    \vspace{-0.2cm}
\end{figure*}
%

% % Comparison of senders 2
\begin{figure}[]
  \centering
  \begin{subfigure}{0.75\columnwidth}
    {
      \includegraphics[width=\linewidth]{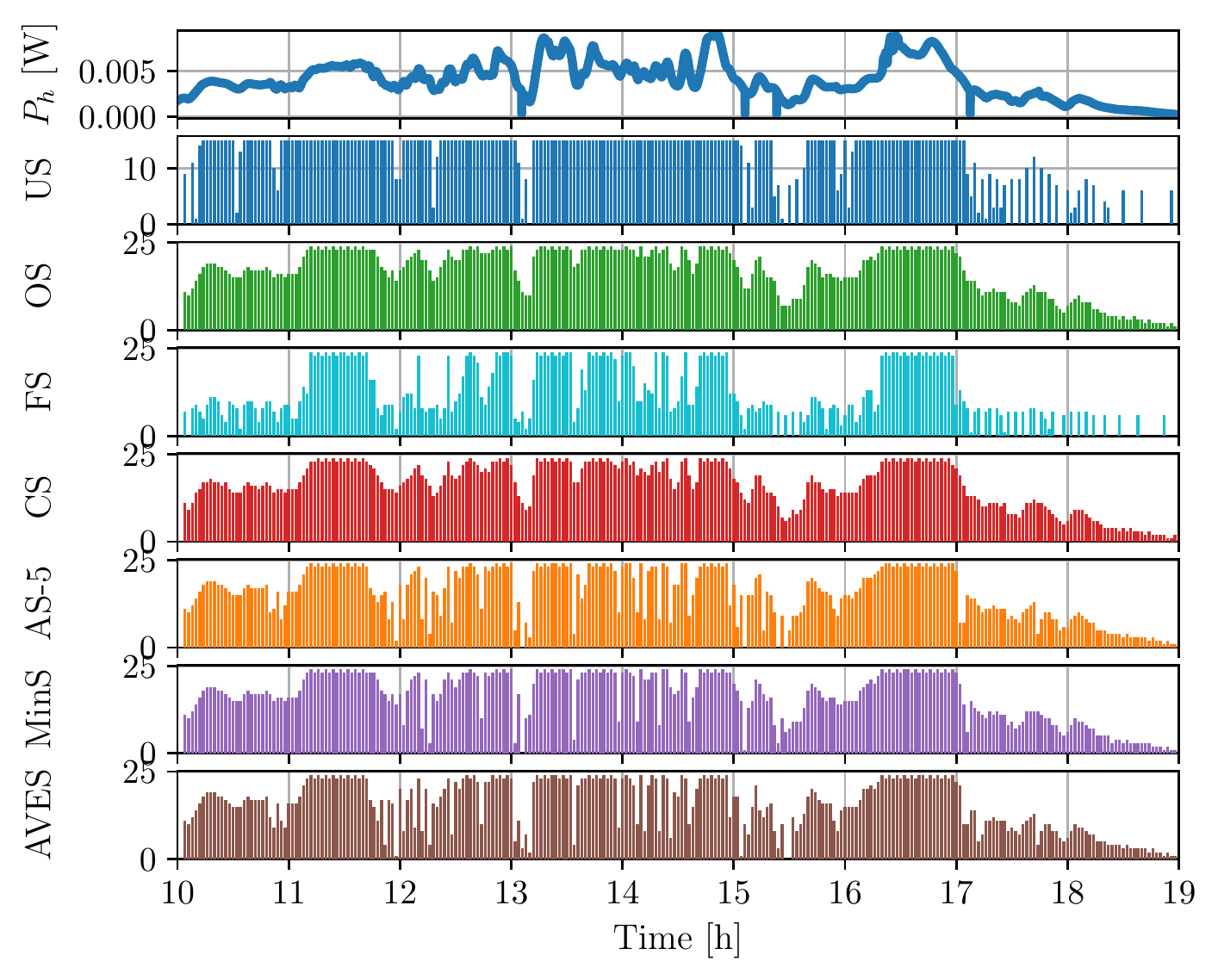}
    }
    \vspace{-0.5cm}
    \caption{Number of transmitted packets for different algorithms.}
    \label{fig:sendertime20eh1}
  \end{subfigure}
  \hspace{0.7cm}
  \begin{subfigure}{0.6\columnwidth}
    {
      \includegraphics[width=\linewidth]{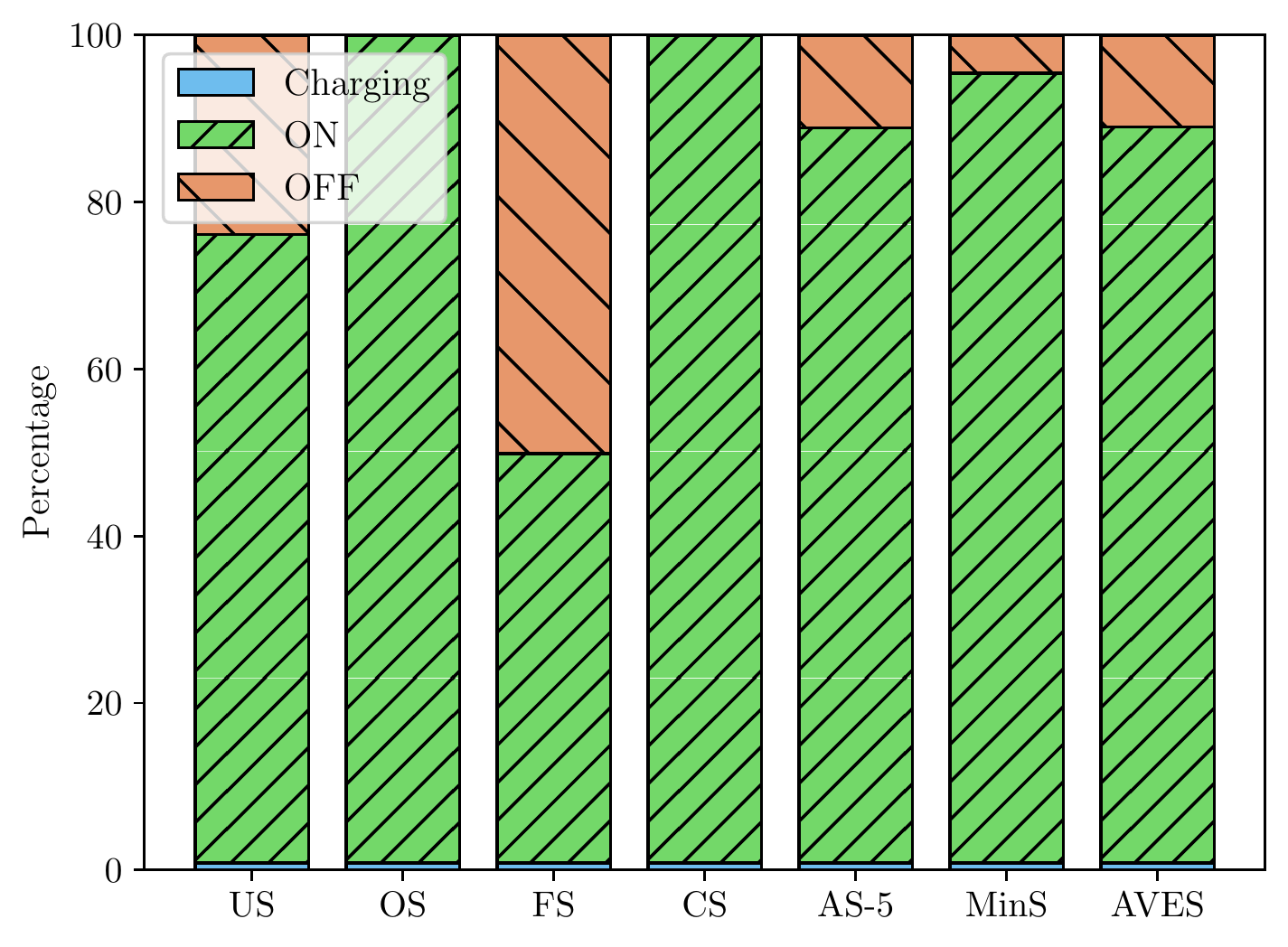}
    }
    \vspace{-0.5cm}
    \caption{Fraction of time spent in different states.}
    \label{fig:onoff20eh1}
  \end{subfigure}
  \caption{Comparison of scheduling behaviors (C=20~mF, trace B).}
  \label{fig:senderstimeEh-2}
   \vspace{-0.5cm}
\end{figure}

\subsubsection{Comparison of scheduling approaches}

\begin{figure*}[t]
  \centering
  \begin{subfigure}{0.75\columnwidth}
    {
      \centering
      \includegraphics[width=\linewidth]{packetsCeh-1.pdf}
      \caption{Trace A.}
      \label{fig:packetsC-1}
    }
  \end{subfigure}
  \hspace{0.7cm}
  \begin{subfigure}{0.75\columnwidth}
    {
      \centering
      \includegraphics[width=\linewidth]{packetsCeh-2.pdf}
      \caption{Trace B.}
      \label{fig:packetsC-2}
    }
  \end{subfigure}
  \begin{subfigure}{0.75\columnwidth}
    {
      \centering
      \includegraphics[width=\linewidth]{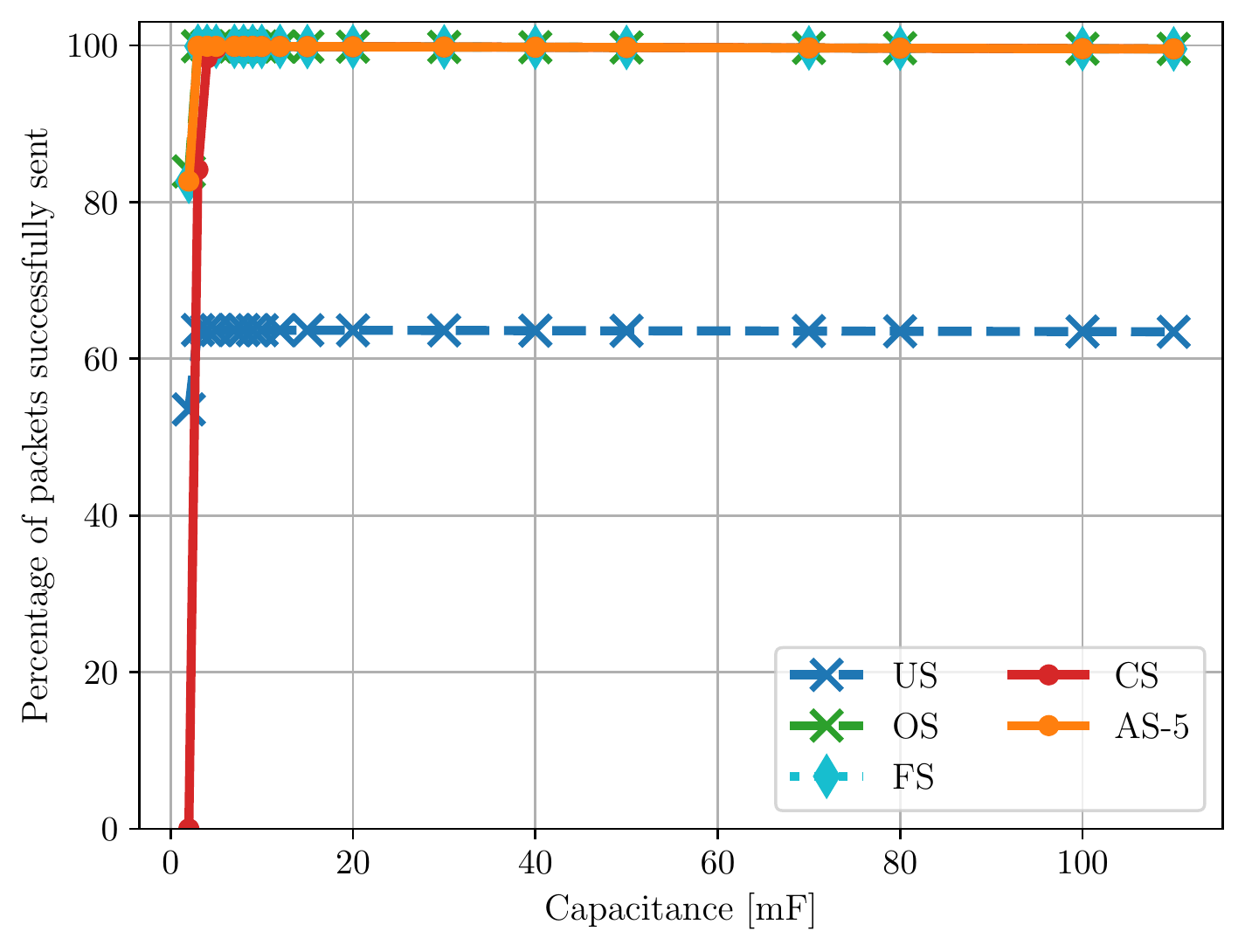}
      \caption{Trace C.}
      \label{fig:packetsC-3}
    }
  \end{subfigure}
  \hspace{0.7cm}
  \begin{subfigure}{0.75\columnwidth}
    {
      \centering
      \includegraphics[width=\linewidth]{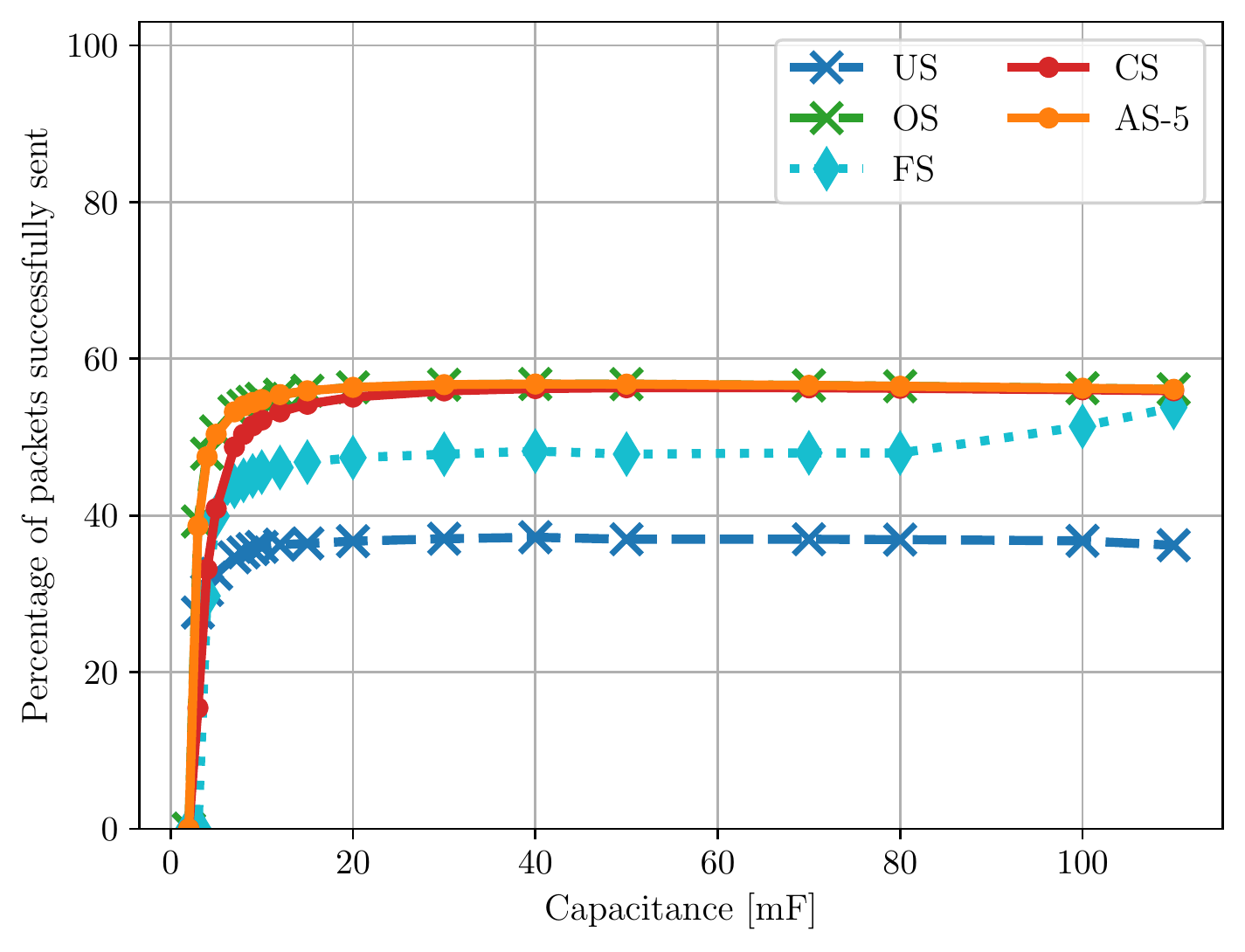}
      \caption{Trace D.}
      \label{fig:packetsC-4}
    }
  \end{subfigure}
  \caption{Comparison of scheduling behavior in terms of percentage of packets successfully sent packets $\epsilon$, for different capacitances and power harvesting traces, PL=5~B.}
  \label{fig:senderscap}
  \vspace{-0.4cm}
\end{figure*}

After showing the potential of energy-aware policies over energy unaware ones,
we now look for the best energy-aware scheduling approach, evaluating different settings. In Fig.~\ref{fig:senderstime} and Fig.~\ref{fig:senderstimeEh-2} we report the number of packets transmitted by the ED over time, and the fraction of time the device spends in the different states (ON/OFF/Charging) for different traces of harvested power.
More in detail, the upper plots in Fig.~\ref{fig:senderstime} show the evolution
of the harvested power for trace A over time, and the number of packets sent by the device in this scenario, with each bar representing the aggregated number of packets sent during two-minute intervals for different schedulers and capacitor
sizes. First, we can notice a correlation between the number of transmitted packets and the harvested power, with both of them increasing between 15:30 and 18:30, when the device was in direct sunlight. Indeed, a higher harvested power
allows the \gls{ed} to charge the capacitor faster, thus maintaining a voltage above the threshold set for transmissions (energy-aware approaches), or preventing the switching off. Secondly, we compare the performance for two capacitance values, i.e., C=20~mF and C=100~mF. As expected, the larger the
capacitor, the longer it takes to charge, as it can be seen comparing Fig.s~\ref{fig:onoff20},~\ref{fig:onoff100} and observing the number of packets sent over time. Indeed, in Fig.~\ref{fig:sendertime100}, for each scheduling algorithm, there is an initial part of the simulation where no packets are sent. 
Then, when the \gls{ed} reaches $V_{th\_high}$, there is a spike of sent packets, since the high voltage (3~V) reached is above the threshold set by the sender, and makes it possible to transmit more packets in a short time. After
this initial transitory phase, from all the graphs in Fig.~\ref{fig:senderstime}, we can see that CS, MinS and the optimal OS algorithms transmit packets rather homogeneously in time for any capacitor size, while the US and FS scheduling algorithms make the device switch off rather often, particularly when the energy harvesting rate is low. 
In particular, US causes periodic switching off of the device because it transmits regardless of the energy level; however, when the capacitor is large, it can keep the device
operational for a longer period, during which it can transmit more than possible for a small capacitance. However, also OFF periods are longer, which results in the larger gaps between transmissions for C=100~mF than for C=20~mF. Similar
considerations hold for FS. In contrast, schedulers that estimate the harvested power (CS, AS, MinS, AVES) achieve better performance, and keep the device operational for most of the time. In particular, the \gls{ed} never switches to the OFF state when considering the CS policy, while this happens in the other approaches for small capacitor sizes (see Fig.~\ref{fig:onoff20}, ~\ref{fig:onoff20eh1}). Note also that the actual performance of the MinS, AS-5 and AVES senders depends on the variability of the harvesting trace: when considering trace B and 20~mF (see Fig.~\ref{fig:senderstimeEh-2}), they are less effective in following the dynamics of the harvesting process and this results in more switching off events. 
This does not necessarily result in a lower number of transmitted packets, as it can be seen in Fig.~\ref{fig:packetsC-2}, because of the more conservative threshold computed by CS, as discussed in the following.
% \textcolor{orange}{why this does not result in a smaller number of packets?
%   because conservative is too conservative and even if it does not switch to off
%   it sends less packets. ADD zoom of trace B, C=20}

%  % Number of packets for different senders and harvesting traces

\subsubsection{Performance evaluation}
\label{sec:perfEval}
In Fig.~\ref{fig:senderscap} we explore the joint impact of capacitor size and harvested power (different traces) on the percentage of packets that are successfully sent by the different scheduling approaches. To reduce clutter and ease the interpretation of the results, we represent the performance of only some of the energy-modeling scheduling algorithms. AVES performed similarly to the AS-5 approach, including the slight drop in performance with trace B, and MinS performed almost equivalently to CS. We only plotted the curves for the CS and AS-5 approaches, which are easier to implement. 
First, we can notice that, for all energy-harvesting traces, the efficiency $\epsilon$ increases with the capacitance up to values around 40~mF, depending on the harvesting rate. Indeed, for trace C,
which has the highest mean harvested power (see Tab.~\ref{tab:power}),  $\epsilon$ reaches 100\%, while in the other scenarios it does not exceed 75\%. The variability of $P_h$ also has an impact. Trace A has higher mean $P_h$, but also higher variability, as it was observed in
Fig.~\ref{fig:senderstime}. Indeed, periods with low $P_h$ correspond to a low amount of transmitted packets, so the performance obtained with trace A is lower than that obtained with trace B.
From Fig.~\ref{fig:senderscap} we can also appreciate the difference between the scheduling approaches. US always performs the worst, transmitting from 20\% to 40\% fewer packets than the OS. The reason is twofold: on the one hand, the \gls{ed} switches off more often; on the other hand, the scheduler drops packets that cannot be transmitted because of \gls{dc} constraints, waiting $I$ more seconds before generating a new packet, as will be better investigated below.
The FS scheduler instead, transmits around 10\% less packets with respect to OS.
FS also transmits more packets for high capacitance values, due to the fact that the capacitor will discharge slowly, staying above $V_{th\_FS}$ (1.82~V) for a longer time and without falling below $V_{th\_low}$.
CS and AS-5, instead, are able to achieve almost optimal performance because, with a proper setting of the threshold value, the \gls{ed} seldom or never switches off. Notice also that, for very small capacitors, no packets are transmitted with the CS algorithm because it sets a higher voltage threshold that is not achievable by a device equipped with only a small 2~mF capacitor assuming no harvesting during the cycle (as conservatively done by CS). AS-5 performs similarly to CS, except for trace B, where the harvested power
changes more frequently over time, leading to incorrect predictions of the harvested power. Furthermore, we can note that for small capacitor sizes (up to 18~mF), AS-5 is able to transmit slightly more packets than CS, although it may cause the \gls{ed} to switch off (as we showed in Fig.~\ref{fig:packetsC-2}). In this case, even if some time is spent in the OFF state recovering energy, the wiser policy of AS-5 allows for the transmission of those packets that, instead, are not sent by CS because of the higher threshold it sets by  conservatively assuming no harvesting during the transmission cycle. 

% % Comparison of SMA senders
\begin{figure*}[t]
  \centering
  \begin{subfigure}{0.75\columnwidth}
    {
      \includegraphics[width=\linewidth]{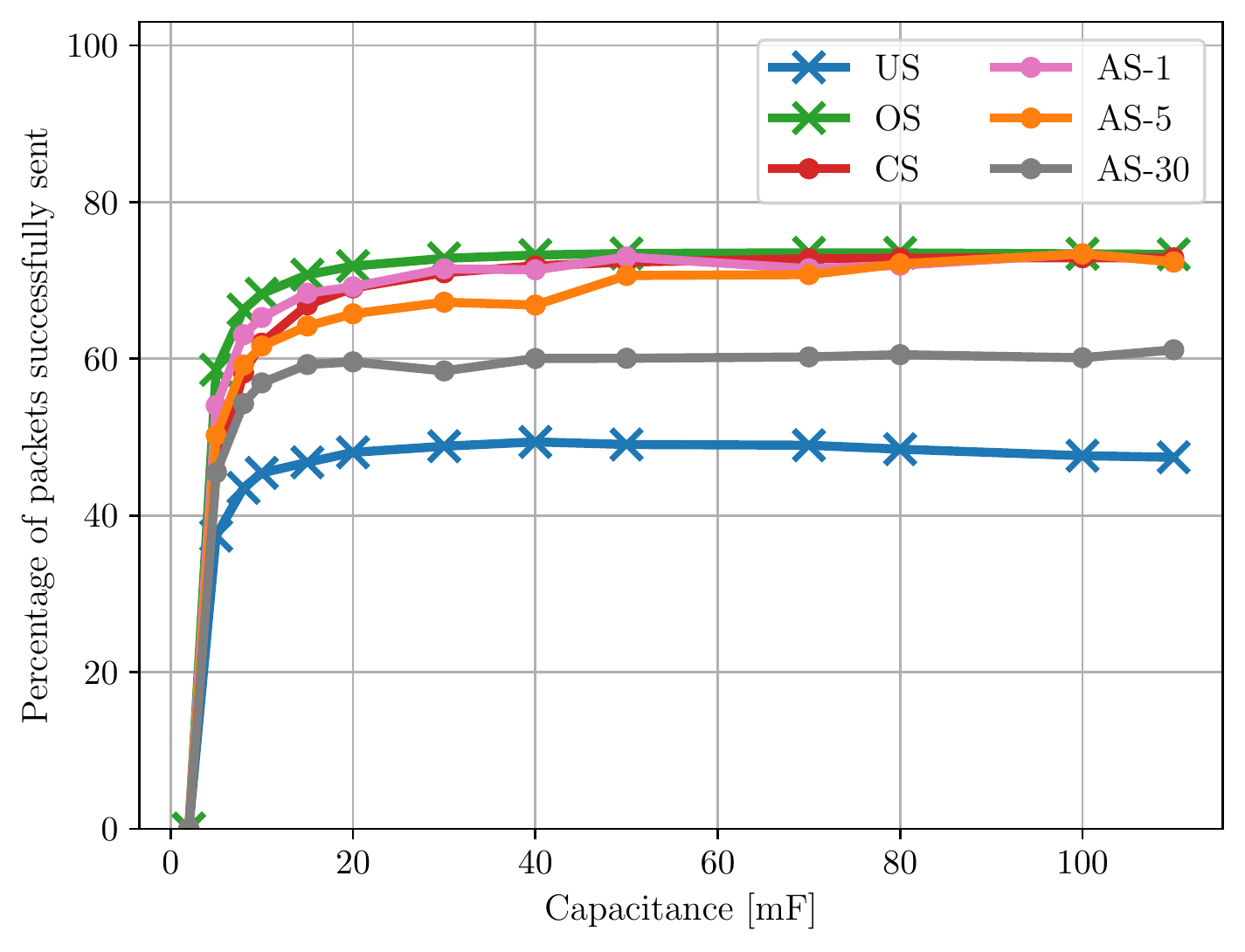}
    }
    \vspace{-0.4cm}
    \caption{Percentage of transmitted packets for different AS-$x$ algorithms, trace B.}
    \label{}
  \end{subfigure}
  \hspace{0.7cm}
  \begin{subfigure}{0.75\columnwidth}
    {
      \includegraphics[width=\linewidth]{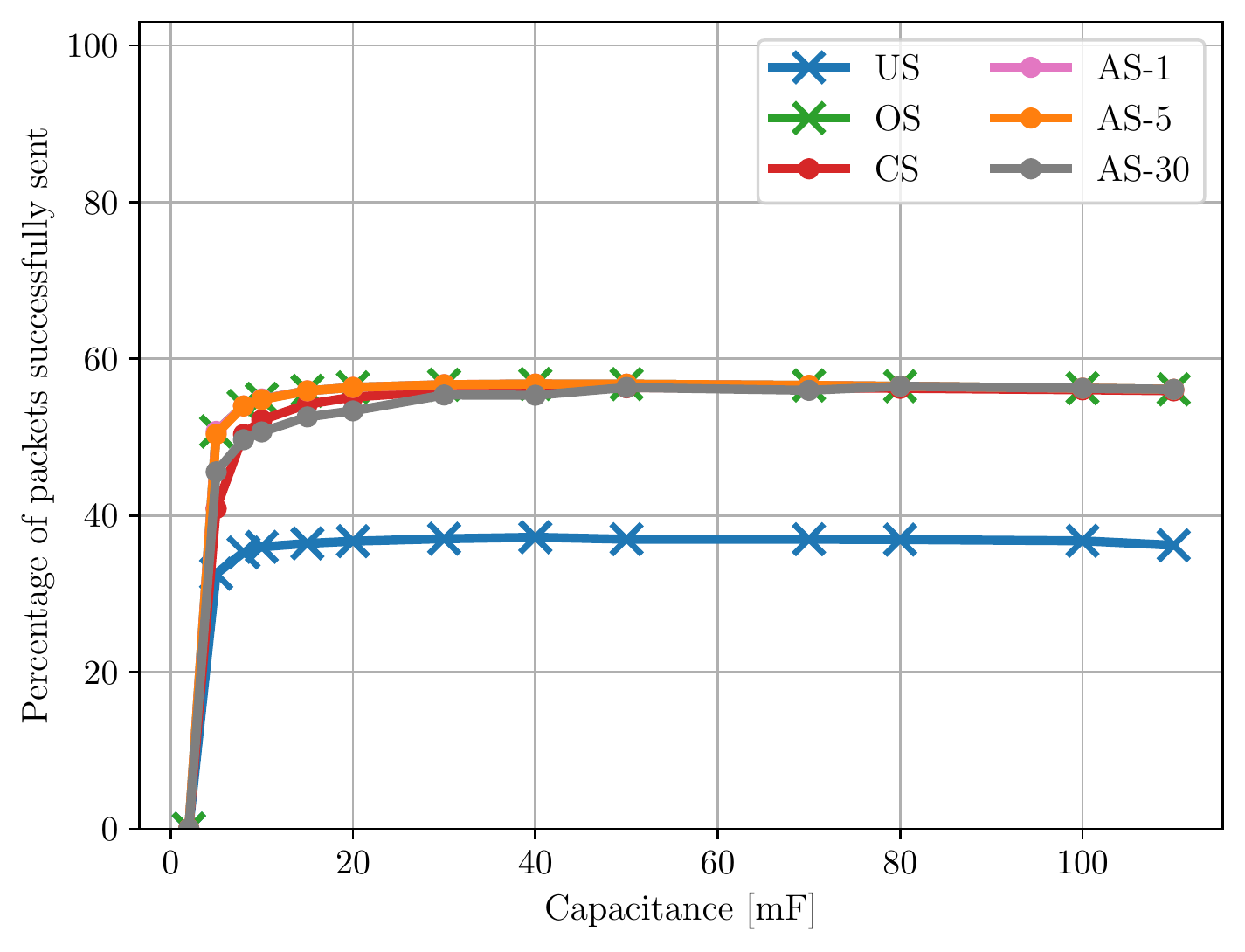}
    }
    \vspace{-0.4cm}
    \caption{Percentage of transmitted packets for different AS-$x$ algorithms, trace D.}
    \label{}
  \end{subfigure}
  \caption{Comparison of AS-$x$ performance with different $x$ values and
    different traces.}
  \label{fig:asComparison}
\end{figure*}

In Fig.~\ref{fig:asComparison} we compare the performance of the AS algorithms taking the average over different time intervals, plotting also the outcomes of US, OS and CS as benchmarks. Note that AS-1 
assumes the current value of the harvesting power will remain constant for the next transmission cycle. Firstly, we can observe that the AS solutions perform differently according to the considered scenario. In particular, there is almost
no difference when employing trace D, which, according to Fig.~\ref{fig:power}, has a smoother behavior over time. In this case, the average harvested power does not change significantly over time intervals of lengths of [1, 30]~s (this may change considering the average over longer periods, but would affect the predictor's capability of following the time evolution of the energy source). On the other hand, when considering trace B, the considered time interval does play a role, and the best solution is to take the average over short periods. This solution also requires less computational and memory resources, which is an advantage when designing ultra-low-power \gls{iot} nodes.

% % Impact of packet size and DC
\begin{figure*}[t!]
  \centering
  \begin{subfigure}{0.75\columnwidth}
    {
      \centering
      \includegraphics[width=\linewidth]{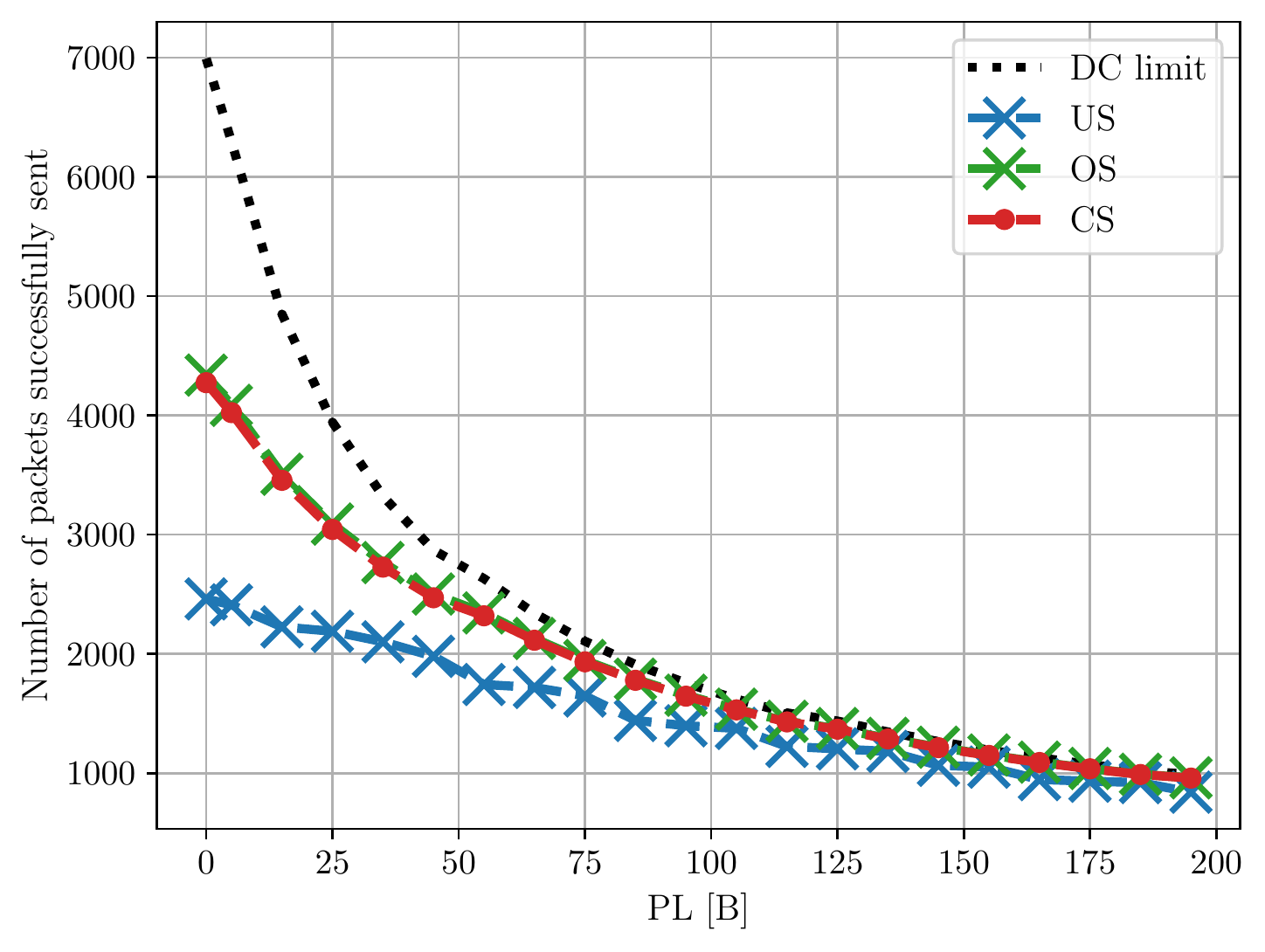}
      \caption{Number of packets transmitted, unconfirmed traffic.}
      \label{fig:psizeunconf}
    }
  \end{subfigure}
  \hspace{0.7cm}
  \begin{subfigure}{0.75\columnwidth}
    {
      \centering
      \includegraphics[width=\linewidth]{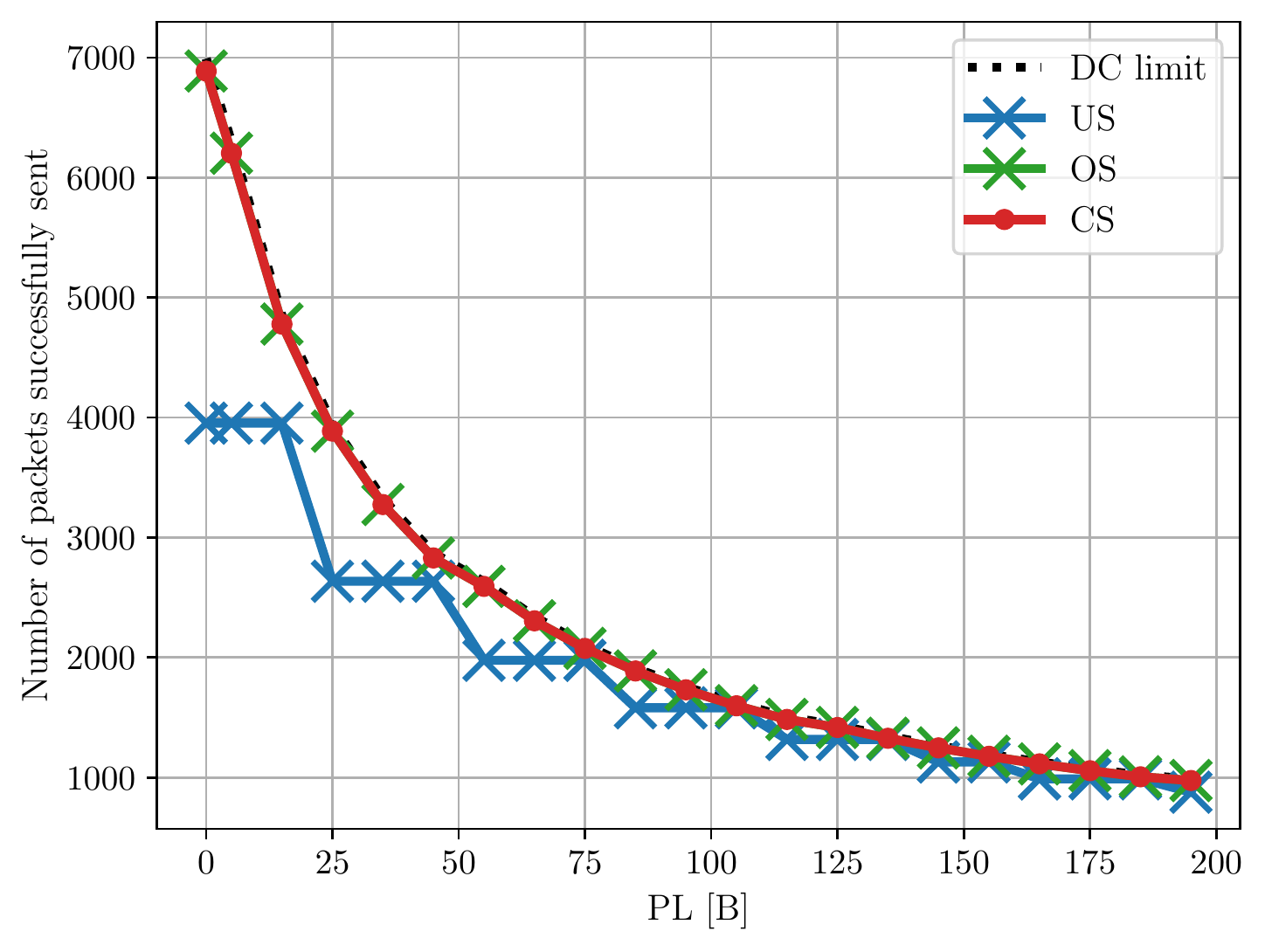}
      \caption{Number of packets transmitted, confirmed traffic.}
      \label{fig:psizeconf}
    }
  \end{subfigure}
  \begin{subfigure}{0.75\columnwidth}
    {
      \centering
      \includegraphics[width=\linewidth]{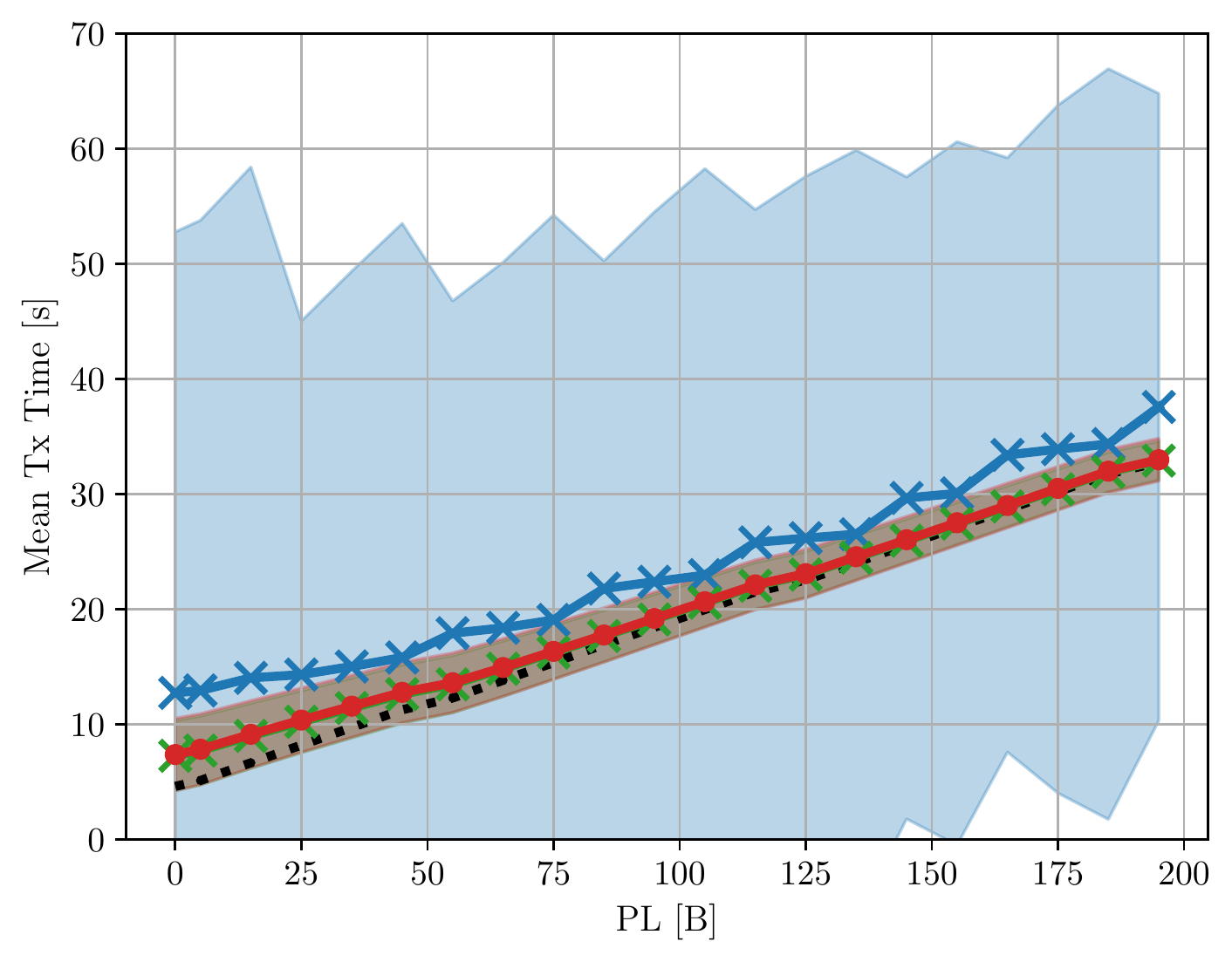}
      \caption{Mean transmission time, unconfirmed traffic.}
      \label{fig:meantxtimeunconf}
    }
  \end{subfigure}
  \hspace{0.7cm}
  \begin{subfigure}{0.75\columnwidth}
    {
      \centering
      \includegraphics[width=\linewidth]{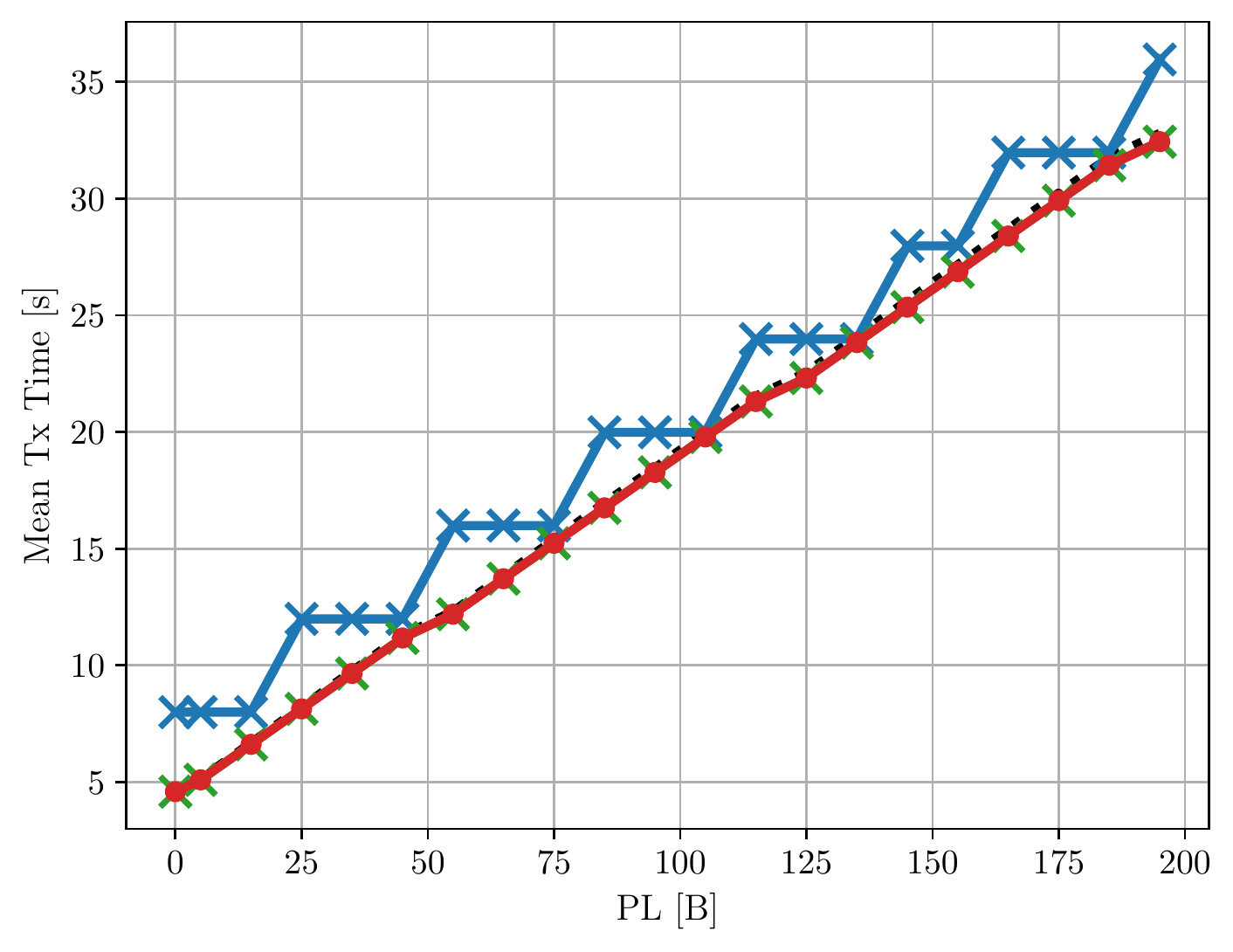}
      \caption{Mean transmission time, confirmed traffic.}
      \label{fig:meantxtimeconf}
    }
  \end{subfigure}
  \caption{Performance comparison for different PL sizes, using trace A, C=40~mF.}
  \label{fig:dcComp}
\end{figure*}
Finally, we explore the impact of the PL and, implicitly, of the \gls{dc} limitations, on the device capabilities, for unconfirmed
(Figs.~\ref{fig:psizeunconf},~\ref{fig:meantxtimeunconf}) and confirmed traffic (Figs.~\ref{fig:psizeconf},~\ref{fig:meantxtimeconf}). According to the results previously shown, the CS approach, although simple, is able to provide performance comparable to OS, performing the best in all the scenarios. Thus, we
now limit our analysis to this scheduler, comparing it to the benchmark solutions and considering a capacitance of 40~mF, which, according to the previous discussion, was the minimum value to obtain good performance with all 4 traces. In Fig.~\ref{fig:dcComp} we compare the benchmark approaches (US, OS) with the CS solution, plotting also the performance bounds imposed by \gls{dc} limitations. From Fig.~\ref{fig:psizeunconf}, we can see that the number of
successfully sent packets decreases for increasing values of PL, with all the solutions converging to the limit imposed by the \gls{dc}. Note that, due to the
limited availability of harvested energy, also the OS approach is not able to reach the \gls{dc} limit. Interestingly, the best performance is obtained when transmitting confirmed packets as
Fig.~\ref{fig:psizeconf} shows. This is somehow counter intuitive, since the \gls{ack}
reception consumes energy. However, a successful \gls{ack} reception in
\gls{rx1} prevents the opening of \gls{rx2}, thus reducing the energy
expenditure and allowing more packets to be transmitted. We can also notice
that, when employing confirmed traffic and US (Fig.~\ref{fig:psizeconf}), curves
have a step shape. This happens because US blindly transmits packets with a
fixed period. Therefore, if the packet transmission is prevented by \gls{dc}
constraints, the following transmission opportunity will occur $I$ seconds later. This is visible, in particular, by inspecting the mean transmission time (see Fig.~\ref{fig:meantxtimeconf}), and it yields the step-shaped curve 
of the number of sent packets. Note that this behavior is not clearly visible
for US with unconfirmed traffic because of the higher mean transmission time.
Also, in Fig.~\ref{fig:meantxtimeunconf} the colored regions represent the
variability of the mean transmission time: it is apparent that, while for CS the
packets are transmitted quite regularly, with US there is no clear transmission
pattern, and intervals between transmissions can be sometimes very long due to
the device switching off, as already observed in Fig.~\ref{fig:senderstime}.
Instead, confirmed traffic yields better results for both metrics, and makes it
possible to have a mean inter-transmission time closer to the \gls{dc} limit,
with negligible variance, in the order of tens of milliseconds, with no
switching off of the device even with the US approach. %  Thus, the energy that is
% not spent for the opening of \gls{rx2} can instead be employed in more frequent
% transmissions, but this may not hold true for other settings (longer packets or
% higher \gls{sf}) and worse environmental conditions (lower harvestable energy).
Furthermore, we should also consider that, although the \gls{dc} regulation
limits the throughput performance, by spacing transmissions apart it allows the
device to stay in sleep mode, charging the capacitor with minimum current
consumption.

These results confirm the benefits of using confirmed traffic to reduce the
energy consumption, as already observed in other literature works, such
as~\cite{delgado2020battery}. However, for high traffic loads, \gls{ack}
transmission may be prevented by the limited \gls{gw}
capabilities~\cite{magrin2019thorough}, and the use of confirmed traffic may
produce packet re-transmissions, causing a dramatic drop on the devices' energy
level, and in the system performance. Therefore, we also tested a solution
where, instead of using confirmed traffic, the sender transmits unconfirmed
packets, but \gls{rx2} is set to use the same \gls{sf} as \gls{rx1}, i.e.,
\gls{sf}~7 in our simulations. The results showed that this approach achieves
basically the same performance as using confirmed traffic, pointing out that the
long duration of \gls{rx2} is a major cause of energy consumption. This solution
can be easily implemented in real systems by modifying the expected value of
\gls{rx2}'s \gls{sf} through appropriate \gls{ns} commands. Note that, in
general, the \gls{ed} will use the lower \gls{sf} that makes the communication
with the \gls{gw} possible, and the \gls{sf} for \gls{rx2} should be consistent
with that. The usage of a \gls{sf} lower than the default
\gls{sf}~12 for \gls{rx2} represents, thus, an improvement compared to the standard
behavior.

% Finally, we inspected the impact of the \gls{dl} packet size. Notice that, in
% general, the \gls{ed} can not be aware of it, and in the threshold
% computation, when a \gls{dl} message is expected, the \gls{ack} is assumed to
% carry no payload. As observed in Fig.~\ref{fig:dlsize}, for some payload
% lengths of the reply packet (2-9~B), CS outperforms OS, since the conservative
% assumption on no harvested power makes it possible to store additional energy,
% which is then employed for packet reception. However, from the figure, it can
% be noted that using confirmed traffic is beneficial only for \gls{dl} packet
% sizes below 10~B, after which the assumption of a \gls{dl} packet with no
% payload has a significant impact on the estimate of the energy consumption,
% and the usage of unconfirmed traffic is preferable.
% %
% %
% \begin{figure}[t]
%   \centering
%   \includegraphics[width=0.45\linewidth]{Dl.pdf}
%   \caption{Number of successfully sent packets for different
%     sizes of the \gls{dl} packet, using trace A, C=40~mF.}
%   \label{fig:dlsize}
% \end{figure}
% %

\section{Conclusions}
\label{sec:conclusions}
In this work, we considered a battery-less energy-harvesting LoRaWAN node, and
provided simulative results assessing the impact of the environmental scenario
(available energy), and design choices (capacitor size, message payload, sender
application) on the capability of the node to send packets. Although the
availability of harvested power has a major impact on the performance, we
identified that a good behavior could be obtained by employing a capacitor of
C=40~mF with the simple CS approach, which, thanks to a conservative assumption
on the harvested power, prevents the \gls{ed} from switching off, reaching a
performance close to the optimal in any environmental conditions. We showed
that, although larger PL sizes increase the transmission time, the effect on the
performance is mostly limited by the \gls{dc} constraint which, on one hand
bounds the amount of packets that can be transmitted, but, on the other hand,
helps charging the capacitor by imposing silent periods. Finally, energy and
communication performance can be increased with the usage of confirmed traffic
or by controlling the \gls{sf} employed in \gls{rx2}.

Although we focused on a single-\gls{ed} scenario, the analysis can be easily extended to network-size simulations. For example, it can be expected that \glspl{ed} located in spatial proximity will likely have similar power availability, which may lead to correlated transmission patterns, possibly increasing the collision probability. One solution that could be investigated is the usage of different \glspl{sf} which, however, will impact on the device's energy (and, thus, communication) performance. Our future work will therefore consider the analysis of this trade-off in large LoRaWAN networks, and investigate possible improvements and solutions. Finally, we remark that substituting batteries with energy-harvesting approaches can provide benefits to environment-related solutions, where it is possible to harvest energy from the environment (e.g., solar light, wind) or from the monitored process itself (stream water flows, geothermal phenomena).

%

%\begin{acks}
\section*{Acknowledgments}
  \small Part of this research was funded by MIUR (Italian Ministry for
  Education and Research) under the initiative ``Departments of Excellence''
  (Law 232/2016), and by the POR FESR 2014–2020 Work Program of the Veneto
  Region (Action 1.1.4) through the project No. 10288513 titled ``SAFE PLACE.
  Sistemi IoT per ambienti di vita salubri e sicuri''. The work was also
  supported by the Flemish FWO SBO S001521N IoBaLeT (Sustainable Internet of
  Battery-Less Things) project, and by the CERCA program, by the Generalitat de
  Catalunya.
%\end{acks}

%\bibliographystyle{ACM-Reference-Format.bst}
%\bibliography{biblio.bib}

  \bibliographystyle{IEEEtran} \bibliography{biblio}

% \balancecolu
\end{document}